\documentclass{article}
\usepackage{graphicx} 
\usepackage{mathtools}
\usepackage{framed,multirow}
\usepackage{algorithm}
\usepackage{algorithmic}
\usepackage{amssymb}
\usepackage{latexsym}
\usepackage{amsmath}
\usepackage{url}
\usepackage{xcolor}
\usepackage{color}
\usepackage{hyperref}
\usepackage{colortbl}
\definecolor{newcolor}{rgb}{.8,.349,.1}
\usepackage{listings}
\usepackage{rotating} 
\definecolor{codegray}{rgb}{0.5,0.5,0.5}
\definecolor{codeblue}{rgb}{0.13,0.13,1}
\definecolor{codepurple}{rgb}{0.58,0,0.82}
\usepackage{geometry}
\usepackage[symbol]{footmisc}

\lstdefinestyle{mystyle}{
    backgroundcolor=\color{white},
    commentstyle=\color{codegray},
    keywordstyle=\color{codeblue},
    numberstyle=\tiny\color{codegray},
    stringstyle=\color{codepurple},
    basicstyle=\ttfamily\footnotesize,
    breakatwhitespace=false,
    breaklines=true,
    captionpos=b,
    keepspaces=true,
    numbers=left,
    numbersep=5pt,
    showspaces=false,
    showstringspaces=false,
    showtabs=false,
    tabsize=2,
}

\newcommand{\bx}{\mathbf{x}}
\newcommand{\DSC}{L_\textrm{DSC}}

\title{Learning to segment anatomy and lesions from disparately labeled sources in brain MRI}
\author{Meva Himmetoglu \\ \small{Computer Vision Lab, ETH Zürich, Switzerland}\\  \small{meva.himmetoglu@vision.ee.ethz.ch} 
\and Ilja Ciernik \\ \small{University of Zürich, Switzerland} \and Ender Konukoglu \\ \small{Computer Vision Lab, ETH Zürich, Switzerland} \and for the Alzheimer's Disease Neuroimaging Initiative \footnote{Data used in preparation of this article were obtained from the Alzheimer’s Disease Neuroimaging Initiative (ADNI) database (adni.loni.usc.edu). As such, the investigators within the ADNI contributed to the design and implementation of ADNI and/or provided data but did not participate in analysis or writing of this report. A complete listing of ADNI investigators can be found at: http://adni.loni.usc.edu/wp-content/uploads/how\_to\_apply/ADNI\_Acknowledgement\_List.pdf} }
\date{}

\begin{document}

\maketitle
\begin{abstract}
Segmenting healthy tissue structures alongside lesions in brain Magnetic Resonance Images (MRI) remains a challenge for today's algorithms due to lesion-caused disruption of the anatomy and lack of jointly labeled training datasets, where both healthy tissues and lesions are labeled on the same images. 
In this paper, we propose a method that is robust to lesion-caused disruptions and can be trained from disparately labeled training sets, i.e., without requiring jointly labeled samples, to automatically segment both. 
In contrast to prior work, we decouple healthy tissue and lesion segmentation in two paths to leverage multi-sequence acquisitions and merge information with an attention mechanism. 
During inference, an image-specific adaptation reduces adverse influences of lesion regions on healthy tissue predictions. 
During training, the adaptation is taken into account through meta-learning and co-training is used to learn from disparately labeled training images. 
Our model shows an improved performance on several anatomical structures and lesions on a publicly available brain glioblastoma dataset compared to the state-of-the-art segmentation methods. 
\end{abstract}
\section{Introduction}
\label{sec:introduction}
Segmentation of structural brain MRI of patients suffering from brain lesions plays an important role in scientific research, quantitative analysis, follow-ups and treatment planning. 
So far, most of the research focused on accurate lesion segmentation~\cite{KAMNITSAS201761,KAMRAOUI2022102312}, leaving aside the segmentation of the healthy structures in the same brain. 
Meanwhile, it has been reported that quantification of many structures along the lineation of the lesion as well as farther areas may play an important role in diagnosis, assessing progression and treatment effects, by providing additional indicators from healthy structures~\cite{PMID:23153407,klein_effect_2002,RASCHKE202299}. 
For example, atrophy is seen as a clinically relevant component of disease progression in multiple sclerosis~\cite{BERMEL2006158} and may be related to cognitive impairment observed post-radiotherapy in brain tumor patients~\cite{RASCHKE202299}. 
Despite the remarkable performance of advanced algorithms for healthy tissue segmentation in lesion-free brain MRI, their performance degrades when applied on patient images with lesions na\"ively because lesions can change the appearance and introduce a domain shift for trained models~\cite{mrbrains13}. 

A straightforward approach of tackling segmentation of lesions and structures simultaneously is with supervised learning~\cite{moeskops2018evaluation}.
However, this approach requires labeling healthy structures of interest in patient images, and for different types of lesions. 
This is clearly expensive and time consuming.
It is also most likely the reason why publicly available datasets containing both healthy structure and lesion annotations on the same images are very rare. 
Furthermore, the supervised approach is limited with the structure set used during labeling. Adding new healthy structures would require relabeling the lesion images. 

An alternative, and arguably more resourceful approach is to use the available datasets focusing on each of the tasks separately: learning healthy tissue segmentation from the publicly available volunteer-based datasets, where healthy tissue labels in large numbers can be obtained with available models designed for this task (~\cite{hcp, mrbrains13}), and lesion segmentation from patient data sets, where only lesion labels are available (~\cite{brats1, brats2, brats3}).
Models that can jointly segment lesion and healthy tissue on the same brain and  that could be trained from such \emph{disparately labeled} task-specific datasets can be tremendously useful. 
Moreover, for new lesion types, generating only lesion labels for a set of images would suffice to train such a joint segmentation model. 

A few recent attempts aimed to solve the joint segmentation problem using disparately labeled datasets. 
The proposed methods leveraged: (i) the availability of multi-sequence acquisitions, dominantly employed in patient imaging, and (ii) the presence of shared sequences in most volunteer and patient datasets, at the least T1w. 
In~\cite{samseg-lesion}, authors define a Bayesian model and solve the resulting optimization problem to segment both healthy structures and lesions in multiple contrasts and resolutions. 
In~\cite{billot_synthseg_2021,billot_joint_2021}, authors achieve robustness for wild variations in images through extensive domain randomization in training. 
Authors in~\cite{dorent}, use shared T1w sequences in healthy and patient images, and assume lesions would not affect the segmentation predictions done based on the T1w sequence. 
Evidence given in their articles, as well as our initial experiments, suggest that this is indeed the case for lesions that do not grossly disrupt the brain anatomy, such as white matter hyper-intensity and multiple sclerosis lesions. 
However, lesions that cause gross disruptions, such as brain tumors, cannot be handled by this approach because the assumption that the lesion does not affect the segmentation prediction on the T1w sequence no longer holds.  

In this work, we develop a new algorithm to exploit multi-sequence datasets; we use multiple available sequences to extract lesion information and the T1w sequence to extract healthy structure information.
There are two main differences compared to previous work, and in particular to the closer previous attempt in ~\cite{dorent}. 
First, we propose to learn to fuse features extracted from different sequences through an attention mechanism, inspired from~\cite{zhang_modality-aware_2021} that proposed a multi-sequence segmentation approach for liver tumor segmentation from multi modal computer tomography images. 
This minimizes adverse effects of concatenating different sequences during feature extraction process. 
For instance, when using patient and volunteer datasets together, where the former often has more sequences than the latter, our approach removes the need to synthesize lesion-specific sequences for volunteer samples.
Moreover, this strategy allows developing separate paths for tissue and lesion segmentation, each pre-trained for the respective tasks, and help attain higher segmentation accuracy for lesions when fused.
Second, we employ image-specific adaptation at inference time to reduce the influence of lesion areas on predictions of healthy tissue segmentation. 
The adaptation takes into account a preliminary lesion segmentation into account to improve robustness of healthy tissue segmentation to lesion-caused disruptions on the anatomy.
The adaptation process is taken into account during training through a meta-learning strategy. 

\section{Related Work}

\noindent\textbf{Multisequence datasets} offer rich information that can be harnessed for more accurate segmentation.
Effective integration of information from these different sequences, such as T1w, FLAIR, contrast enhanced T1w and T2w, is important for the segmentation task, since each sequence can offer complementary information about the anatomy and the pathology.
There are two particularly relevant points we would like to highlight. 
First, imaging studies for brain lesions often include multiple MRI sequences to display the complementary information. 
Today there are publicly available multi-sequence MRI datasets for a number of lesions including lesion segmentation labels.
These datasets enabled development of accurate lesion segmentation models using multiple sequences \cite{nnunet, wangtumor} and even when some sequences are missing \cite{d2net}, \cite{rfnet}. 
Second, healthy structure segmentation in brain MRI today is mostly based on the T1w sequence, in which the contrast between gray and white matter is strong. 
There are publicly available volunteer and neurodegeneration-focused datasets, which have large number of T1w images, cover a wide age-span and are acquired from individuals without gross lesions. 
Well established analysis toolboxes, such as Freesurfer \cite{freesurfer}, SPM \cite{spm_book} and FSL \cite{jenkinson_fsl_2012, woolrich2009bayesian, smith2004advances}, and deep learning-based segmentation algorithms are able to segment lesion-free T1w sequences accurately. 
This capability is even extended to other sequences and varying imaging protocols in \cite{billot_synthseg_2021}.
Importantly, the main sequence for healthy structure segmentation, the T1w sequence, is also almost always present in imaging studies for lesions. 
This common use of the T1w sequence and presence of methods that can accurate segment lesion-free T1w sequences allow tackling the joint segmentation problem through models that can be trained using  disparately labeled datasets. 

\noindent\textbf{Joint segmentation} in this work refers to the simultaneous segmentation of healthy tissue structures and lesions in a brain MRI.
One of the initial works, \cite{moeskops2018evaluation}, evaluated supervised learning approach, for which training samples with both healthy tissue and lesion labels were required. 
As we mentioned earlier, this approach has limitations both in terms of costs of training set generation and in terms of generality. 
To tackle the joint segmentation problem without requiring exhaustively labeled lesion images, recent studies proposed various strategies to jointly train models on disparately labeled datasets. 
\cite{samseg-lesion} extended their previous whole brain segmentation model, which was proposed in \cite{samseg}, for it to handle white matter hyperintensities.
They added an additional lesion label to the probabilistic generative model and model lesion shapes using a variational autoencoder.
In a similar approach, \cite{billot_joint_2021} extended their previous domain randomization based segmentation method to include white matter hyperintensities.
\cite{dorent} took a different approach and utilized lesion free volunteer dataset and domain adaptation techniques, such as data augmentation, adversarial learning and pseudo-healthy generation, to jointly segment images bearing multiple sclerosis lesions and gliomas.
They used feature channel averaging across multiple sequences for images bearing lesions, and used T1w images for lesion-free images.
\cite{LIU2024110636} have recently proposed a model to segment lesions and vessels jointly on fundus images.
They introduce an adversarial self training strategy and a cross task attention module to improve performances of each individual task.

\section{Methodology}
\subsection{Problem Formulation}\label{sec:problem_formulation}
The aim of our model is to obtain joint segmentation of tissues and lesions on images bearing lesions.
We have a particular focus on lesions that grossly disrupts anatomy. 
To this end, we illustrate our work on brain tumors, and more specifically gliomas. 
As we mentioned, large scale imaging datasets of patients with lesion and anatomy segmentations are not widely available for training joint segmentation models in a fully supervised fashion.
Therefore, we are using two disparately labeled datasets curated for different tasks.
First is composed of lesion-free images and segmentations of anatomical structures, 
and the second is composed of images of patients with lesions and segmentation of the lesions, but no anatomical structure segmentation.
We assume patients have multi-sequence MRI available and there is at least one overlapping sequence between the lesion-free and patient datasets, which in our case is the T1w sequence. 

We describe the proposed method starting from the inference stage. 
We denote an imaging series with $\bx$. 
Imaging series, especially for patients with lesions, include multiple sequences, thus $\bx$ denotes a set of images. 
In this work, without loss of generality, we assume there are two sequences $\bx = \{x^{T_1}, x^F\}$, T1w and FLAIR, respectively. 
Each image has $N$ voxels and we also assume that the different sequences in $\bx$ are co-registered, which can be easily attained using a rigid registration in most cases. 
Even though we assume two sequences, more can be considered with minimal modifications to the proposed model. 

The ultimate goal of a joint segmentation model is to predict a voxel-wise segmentation map for $\bx$, which we denote as $\hat{y}$ and its ground truth as $y$.
$\hat{y}$ includes labels for the lesion and a set of healthy structures. 
In this work we use \{\emph{gray matter, white matter, basal ganglia, ventricles, cerebellum, brain stem and the background}\} as the set of healthy structures in our illustrations and experiments. 
We denote the label for the lesion with $c_L$ and labels for $M$ anatomical structures with $\{c_A^0,\dots,c_A^M\}$, where $c_A^0$ refers to the background. 
Hence, each voxel $n$ in $\hat{y}$ can be one and only one of these structures, i.e., $\hat{y}_n \in \{c_A^0,\dots,c_A^M, c_L\}$. 

\begin{figure}[!h]
    \centering
    \includegraphics[width=0.5\textwidth]{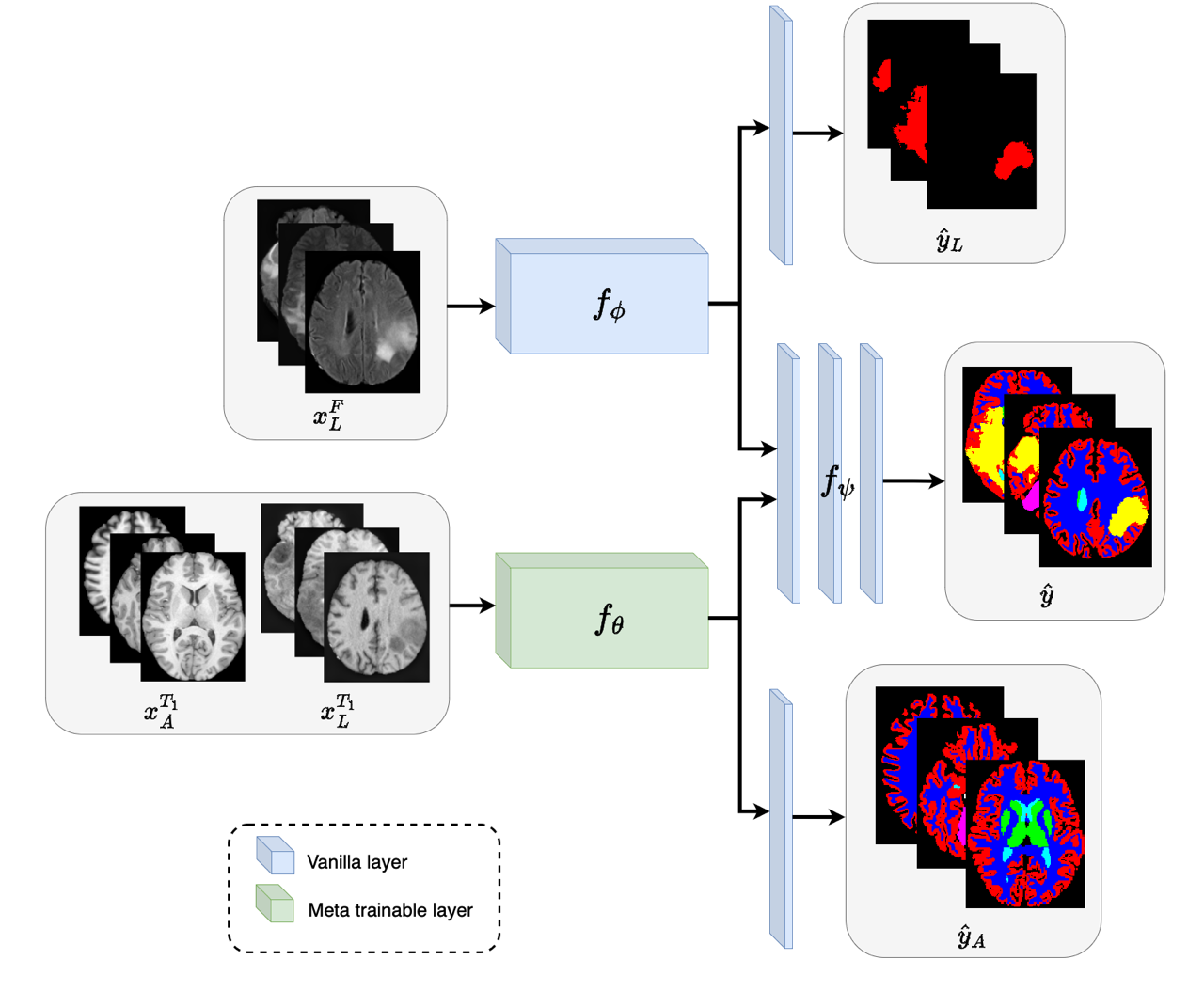}
    \caption{Model overview. U-shaped feature extractors $f_\theta$ and $f_\phi$ takes images with lesion ($x_L$) and lesion free images ($x_A$) images and output segmentations. For joint segmentations ($\hat{y}$), both modalities' feature extractors are used. In the proposed pipeline, first $f_\theta$ is trained with lesion free datasets with tissue labels; and $f_\phi$ is trained with patient datasets and lesion labels to segment task specific datasets. Second, $f_\theta$  is trained in a meta-fashion such that the T1w features can be adapted to presence of lesions in test time. Lastly, $f_\psi$ is added in training such that FLAIR and T1w features can be fused into a joint segmentation output ($\hat{y}$) for images bearing tumors.}
    \label{fig:model}
\end{figure}

We denote our joint segmentation model with $f(\bx; \phi, \theta, \psi)$, which is illustrated in Figure~\ref{fig:model}. 
$f(\bx; \phi, \theta, \psi)$ is a 3D neural network architecture and it is formed of three components, i.e., $f(\bx; \phi, \theta, \psi) \triangleq f_{\psi}(f_{\theta}(x^{T_1}), f_{\phi}(x^F)) = f_{\psi}(w^{T_1}, w^F)$. 
$f_{\theta}$ and $f_{\phi}$ extract features from the T1w and FLAIR sequences, respectively. 
$f_{\theta}$ is responsible for generating the feature map $w^{T_1}$ that is useful for segmenting healthy structures and $f_{\phi}$ is responsible for generating the feature map $w^F$ that is useful for segmenting the lesion. 
$f_{\psi}$ fuses these feature maps through an attention mechanism to predict the joint segmentation map.
This separation of responsibilities and the attention-based fusion is critical as it allows training this model with disparately labeled datasets, which we further detail in Section \ref{sec:training}. 
\footnote{
When more sequences are used, either further components can be added to the model or each sub-model can be modified to use more sequences. 
Since more sequences are often used for lesion characterization, we envision $f_{\phi}$ to be modified to take into account further sequences. }
$f(\bx; \phi, \theta, \psi)$ outputs per voxel probabilities as common in deep learning-based segmentation models. 
For a  voxel $n$ the model outputs the probability of that voxel belonging to a class $c$, i.e., $[f(\bx; \phi, \theta, \psi)]_{c,n} = [p(c|\bx)]_n$, where $c\in \{c_A^0,\dots,c_A^M, c_L\}$. 
Then the final maximum-likelihood segmentation is $\hat{y}_n = \arg_{c}\max [p(c|\bx)]_n$. 

While the joint model $f(\bx; \phi, \theta, \psi)$ generates the joint segmentation $\hat{y}$, information coming from the different components $f_{\theta}$ and $f_{\phi}$ are used for two more tasks. 
First, $f_{\theta}(x^{T_1})$ is used by a shallow network, $g_A(\cdot)$, to generate a healthy structure-only segmentation map, $\hat{y}_A$, with $\hat{y}_{A,n} \in \{c_A^0, \dots, c_A^M\}$ for each voxel $n$.
Second, $f_{\phi}(x^F)$ is used by another shallow network, $g_L(\cdot)$, to generate a binary lesion segmentation map, $\hat{y}_L$, with $\hat{y}_{L,n} \in \{0,1\}$. 
These tasks are the reasons why we stated that $w^{T_1}$ and $w^F$ are responsible for structure and lesion segmentations, respectively. 

An important component of this model is that at inference time $f_{\theta}(x^{T_1})$ should be able to generate representations that can be used to segment healthy structures. 
In the presence of large disruptive lesions, $x^{T_1}$ can deviate from healthy anatomy and thus $w^{T_1}$ can be affected in ways that makes it not suitable for accurate healthy structure segmentation. 
Furthermore, this effect may be difficult to predict in advance. 
To account for this, the model includes an inference-time adaptation step that aims to alleviate effects of a lesion on $w^{T_1}$. 
We model an image bearing a lesion, $x$, as a combination of two spatial partitions with voxel-wise binary maps $A$ and $L$ as 
\begin{equation}x = x \odot A + x \odot L,\end{equation} where \begin{equation}A,L = \{0,1\}^N \textrm{with } A+L = 1 \textrm{ and } A\odot L = 0,\end{equation}
where $A$ and $L$ corresponds to anatomy and lesion areas respectively, and we denoted the element-wise product, i.e., Hadamard product, with $\odot$. 
Under this model, the effect of the lesion area on $w^{T_1}$ can be more clearly expressed with
\begin{equation}w^{T_1} = f_{\theta}(x^{T_1}) = f_{\theta}(x^{T_1}\odot A + x^{T_1} \odot L).\end{equation}

Ideally, the information required for segmenting healthy structures should not depend on the intensity within the lesion area. However, a model trained with only healthy anatomy would use the information that area assuming it is useful information for its task. We tackle this issue with an inference-time adaptation.
Assume for a moment that $L$ is known for a given image. 
In this case, one way to enforce that $w^{T_1}$ does not rely on the lesion is to minimize its sensitivity to $x^{T_1}\odot L$. 
This can be achieved by enforcing that $w^{T_1}$ does not change under randomization of the content in the lesion area, i.e., 
\begin{equation}f_{\theta}(x^{T_1}\odot A + x^{T_1} \odot L) = f_{\theta}(x^{T_1}\odot A + \tilde{x} \odot L) \end{equation}
or in terms of the features enforcing $w^{T_1} = \tilde{w}^{T_1},$
where $\tilde{x} \sim p_{OOD}$ is a random sample from an out-of-image distribution and $\tilde{w}^{T_1}$ is the feature map under the randomization. 
As the lesions can be large and arbitrarily located, we believe this condition should be enforced during the inference process. 
Since the goal is healthy tissue segmentation, our inference procedure includes an optimization to maximize the consistency between $g_A(w^{T_1})$ and $g_A(\tilde{w}^{T_1})$ for arbitrary $\tilde{x}\sim p_{OOD}$ with respect to $\theta$.
To this end, the Dice's similarity coefficient (DSC) is used, i.e., 
\begin{equation}\min_{\theta} \mathbb{E}\left[\DSC\left( g_A(w^{T_1}), g_A(\tilde{w}^{T_1}) \right)\right],\end{equation} where the expectation is with respect to $\tilde{x}$ and $\DSC$ is the average soft Dice loss \cite{dice,sorensen1948method} across all structures. Ideally, one would have liked to maximize DSC between predictions and ground truth segmentation maps, however, remember that this optimization happens at inference-time, so the model does not have access to ground truth segmentation of the healthy structures. Effectively, the consistency loss is a surrogate loss.

It is important to note that maximizing the consistency alone may lead to divergence, where the predicted healthy structure segmentations $g_A(w^{T_1})$ and $g_A(\tilde{w}^{T_1})$ are consistent but wrong. 
To remedy any possible divergence, we can add supervision through a lesion-free image and its corresponding ground truth label maps, which can be taken from an existing database of volunteers with T1w brain scans. 
More specifically, we can add another supervised loss in the form of $\DSC\left(g_A(f_{\theta}(x_A^{T_1})), y_A\right)$ for a chosen lesion-free T1w image $x_A^{T_1}$ and its corresponding ground truth healthy structure segmentation map $y_A$. Here, we note that the separation in the network makes it possible to use a lesion-free T1w image to avoid divergence without requiring other sequences. 
Neither $g_A(\cdot)$ nor $f_\theta(\cdot)$ requires any sequence other than the T1w. 

The last important point is to address the assumption we made earlier, having a binary $L$ map for a given image. 
This is naturally not possible for test samples, since $L$ roughly corresponds to the lesion segmentation and this is one of the outputs we are seeking. 
So $L$ needs to be estimated from the test sample.
To this end, given a test sample $\mathbf{x}=\{x^{T_1}, x^F\}$, the lesion segmentation prediction is obtained from $\hat{y}_L \triangleq g_L(f_{\phi}(x^F)) = g_L(w^F)$ and used as an estimate of $L$. 
Note that the separation of paths in the network becomes useful, allowing such an estimation.

The resulting loss that is minimized during the inference-time adaptation is then given as 
\begin{align}\label{eqn:innerloss}\min_{\theta }\mathcal{L}_i(\theta;x_A^{T_1},y_A,x^{T_1},\hat{y}_L) = \min_{\theta} &\ \mathbb{E}\left[\DSC\left( g_A(w^{T_1}), g_A(\tilde{w}^{T_1}) \right)\right] + \nonumber  \DSC\left(g_A(f_{\theta}(x_A^{T_1})), y_A\right),\end{align}
where $x_A^{T_1}$ denotes a chosen lesion-free T1w image and $y_A$ denotes its corresponding ground truth label map, $x^{T_1}$ is the T1w image of a given test case that may have a lesion, i.e., $\mathbf{x}=\{x^{T_1}, x^F\}$, $\hat{y}_L$ is the predicted binary lesion segmentation map $g_L(f_{\phi}(x^F))$ for the test case, and the feature maps are given as $w^{T_1}=f_{\theta}(x^{T_1}\odot (1 - \hat{y}_L) + x^{T_1} \odot \hat{y}_L)$ and $\tilde{w}^{T_1} = f_{\theta}(x^{T_1}\odot (1-\hat{y}_L) + \tilde{x} \odot \hat{y}_L)$ with $\tilde{x}\sim p_{OOD}$ being sampled as described above.

\begin{algorithm}
\caption{Inference for Images Bearing Lesions}
\label{alg:inference}
\begin{algorithmic}[1]
\REQUIRE $\bx_L = \{x_L^{T_1}, x_L^F\}$ : image series bearing lesions
\REQUIRE $\bx_A = \{x_A^{T_1}\}$ : lesion free image series
\REQUIRE $f_\theta$ : T1w feature extractor
\REQUIRE $f_\phi$ : FLAIR feature extractor
\REQUIRE $g_A(\cdot)$, $g_L(\cdot)$: segmentation layers
\FOR {$x_L$ in $X_L$}
    \STATE Generate feature maps: 
    
         \quad $w^{T_1} \gets f_{\theta}(x^{T_1})$
         
         \quad $w^F \gets f_{\phi}(x^F)$
        
    \STATE Predict initial segmentations:
    
         \quad $\hat{y}_A \gets g_A(w^{T_1})$
    
         \quad $\hat{y}_L \gets g_L(w^F)$

\STATE Minimize sensitivity over lesion by maximizing consistency

   \quad Generate randomized lesion content: $\tilde{x} \sim p_{OOD}$
   
   \quad  Generate modified feature map: $\tilde{w}^{T_1} \gets f_{\theta}(x^{T_1} \odot (1-\hat{y}_L)+ \tilde{x} \odot \hat{y}_L)$

   \quad  Optimize $\theta$ to maximize consistency:
  
   \quad $\theta' \gets \arg_{\theta}\max \mathbb{E}[\DSC\left(g_A(w^{T_1}), g_A(\tilde{w}^{T_1})\right) + \DSC\left(g_A(f_{\theta}(x_A^{T_1})), y_A\right)]$
   
\STATE Get joint segmentation with new parameters $\theta'$

    \quad $\hat{y}_A \gets g_A(f_\theta'(x^{T_1})$
    
    \quad $\hat{y} \gets   f_\psi(f_\theta'(x^{T_1}), f_\phi(x^F)) $

 \textbf{Return} $\hat{y}$, $\hat{y}_A$ 
\ENDFOR
\end{algorithmic}
\end{algorithm}

The inference procedure described above is summarized in Algorithm \ref{alg:inference}. In the next section, we explain the components and the architecture of the model before going into the training procedure.

\subsection{Model Overview}

Proposed model consists of two branches, one for each sequence of interest, namely $f_\theta$ and $f_\phi$ networks for T1w and FLAIR images, respectively. 
Both branches have U-Net like contracting and expanding structures and output feature maps $w^{T_1}$ and $w^F$, which are converted to segmentations maps through two shallow networks $g_A$ and $g_L$, as shown in Figure \ref{fig:model}. 
In order to leverage the multiple sequences, we construct a fusion part, namely $f_\psi$, at the end of the two branches, taking the last layer features of both as input and outputting a joint segmentation of both anatomical structures and lesions through a few convolutional layers.

The multi-branch architecture with sequence-specific feature extraction and an attention-based fusion of the extracted features was proposed by \cite{zhang_modality-aware_2021} for segmentation of liver tumors.
We use an architecture inspired by this work, to leverage the different appearances of anatomical structures and lesions in different sequences. 
However, different than \cite{zhang_modality-aware_2021} we focus on joint segmentation of tissue and lesion, and train the network with a novel strategy apt to this task. 

\subsection{Training Procedure} \label{sec:training}

The proposed training strategy in summary is as follows. 

(1) Train each task specific branch separately to obtain good segmentations from each. 

(2) Fine-tune the T1w branch with meta-learning using synthetically generated data where ground truth segmentations for both healthy structures and tumor is available.

(3) Use fine-tuned T1w branch to generate pseudo-labels for the healthy structures in images with tumors, for which only tumor segmentation labels are available.

(4) Use the pseudo-labels and ground truth tumor labels to train the joint model, where meta-learning is utilized for the training of the T1w branch this time based on segmentation predictions of the FLAIR branch. 

Below, we describe these steps in detail. Specifically, we describe the training in three episodes: pretraining, meta co-training and joint training. 

As discussed, our aim is to train the model using disparately labeled datasets.
To this end, our training set consists of two labeled datasets $D_A = (X_A, Y_A) = (\{\bx_A\},\{y_A\})$ and $D_L = (X_L, Y_L)=(\{\bx_L\},\{y_L\})$. 
The former is comprised of a set of lesion free images $X_A$ and corresponding healthy structure labels $Y_A$, and the latter is comprised of multi-sequence images bearing lesions $X_L$ and corresponding lesion segmentations $Y_L$. 
We assume that for each subject $\bx_A \in X_A$ is formed of a single T1w image while for each subject $\bx_L \in X_L$ is formed of a T1w and a FLAIR sequence. 
Any $y_A \in Y_A$ is a multi-label map including healthy structure labels and $y_L \in Y_L$ is a binary segmentation map highlighting the lesion area. 
We also do not require any overlap between the datasets, and work on the assumption that they have none. 

\subsubsection{Pretraining}
In the first episode of the proposed pipeline, task-specific feature extractors are trained to obtain segmentations by utilizing labeled datasets. 
We utilize FLAIR images ${x}^F_L $ and lesion labels $ {y}_L $ from the lesion dataset in a supervised fashion with Dice loss.
Similarly we utilize T1w images ${x}_A^{T}$ and anatomical tissue labels $y_A$ from the lesion-free dataset in a supervised fashion trained with Dice loss. These losses lead to the following independent optimizations for the separate paths:

\begin{align}
    &\arg_{g_A, \theta} \max \sum_{(x_A^{T_1}, y_A)\in D_A}\DSC (g_A(f_\theta(x_A^{T_1})), y_A)\\
    &\arg_{g_L, \phi} \max \sum_{(x_L^F, y_L)\in D_L} \DSC (g_L(f_\phi(x_L^F)), y_L).
\end{align}

Two lesion free datasets are used in pretraining of $g_A$ and $f_\theta$, one obtained from healthy volunteer subjects and the other one obtained from elderly subjects from Alzheimer's Disease Neuroimaging Initiative.
For both of these datasets, there are no disruptive lesions visible in T1w images, and we obtained the "ground truth" segmentations with the Freesurfer software \cite{freesurfer}.
Freesurfer segments a large number of cortical and subcortical structures. 
While all these labels could be used, in this work, we focused on a reduced set of labels in our experiments. 
This reduced set is obtained by merging Freesurfer labels to the target labels mentioned above. 
We refer the reader to the Appendix \ref{sec:labeling} for the exact matching function. 
At the end of pretraining stage, we have sequence- and task-specific networks for lesion and anatomical tissue segmentation, namely $f_\phi$, $f_\theta$ and their corresponding shallow networks $g_L$ and $g_A$.

\subsubsection{Meta Co-Training}
\begin{figure}[!h]
    \centering
    \includegraphics[width=0.5\textwidth]{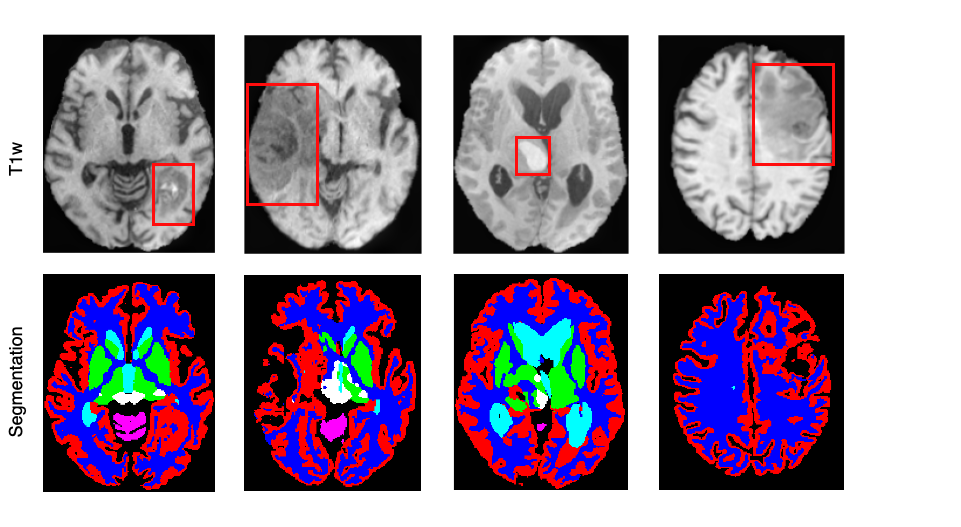}
    \caption{T1 weighted image examples from BraTS dataset (top) and their tissue segmentations with anatomical structure segmentation network after the pretraining stage (bottom). Red boxes highlight areas with lesions. The corresponding parts in the predictions are less accurate than the other parts of the brain (Labels are depicted as follows: red: gray matter, blue: white matter, cyan: ventricle, green: basal ganglia, white: brain stem and magenta: cerebellum).}
    \label{fig:healthyseg}
\end{figure}

After the pretraining stage, T1w path, i.e., $g_A\circ f_\theta$ is optimized to segment anatomical structures in  images without disruptive lesions.
Images bearing large lesions, such as tumors, however, would have imperfect segmentations around the lesion, since the intensity variations caused by the lesion have not been seen in the pre-training. 
Examples of outputs of the anatomical tissue segmentation network for images with tumors can be seen in Figure \ref{fig:healthyseg}.
 
Our method uses the inference time adaptation step described above to reduce the sensitivity of $w^{T_1}$, and therefore the T1w path $f_\theta$ and the resulting healthy tissue segmentation, to the presence of lesions. 
We take this adaptation into account during the training through the meta-learning approach. 
\cite{maml} introduced meta-learning where the training determines parameters that can be easily adapted to new tasks in the few-shot learning scenario. 
We use the same strategy, but to reduce the sensitivity of feature map $w^{T_1}$ to present lesions as described in Section~\ref{sec:problem_formulation}. 
An algorithmic summary of meta training episode is given in Algorithm \ref{alg:meta}. In the algorithm, $x_A$ and $x_P$ are 3D images from lesion free dataset and the generated pseuodo unhealthy dataset respectively.

\begin{algorithm}[t]
\caption{Co-training with Meta Updates}
\label{alg:meta}
\begin{algorithmic}[1]
\REQUIRE $\alpha$, $\beta$: step size hyperparameters
\REQUIRE $\{x_A^{T_1}\}$, $\{y_A\}$: lesion free images and their labels
\REQUIRE $\{x_P^{T_1}\}$, $\{y_P\}$: pseudo-lesioned images and their labels
\STATE get pretrained tissue segmentation model parameters $f_\theta$
\WHILE{not done}
\STATE Sample batch of images from $x_A$ and $x_P$
 \STATE Calculate $\mathcal{L}_i (\theta; x_A^{T_1}, y_A, x_P^{T_1}, y_P)$
 \STATE Compute adapted parameters with gradient descent: $\theta'=\theta-\alpha \nabla_\theta \mathcal{L}_i$
 \STATE Calculate $\mathcal{L}_o ( {\theta'}, x_A, y_A, x_P, y_P)$
 \STATE Update $\theta \leftarrow \theta - \beta \nabla_\theta \sum  \mathcal{L}_o$
\ENDWHILE
\end{algorithmic}
\end{algorithm}

To train the T1w path using meta-learning, we start from the pre-trained weights $\theta$. 
The aim of the meta learning training is to determine new $\theta$, which when updated at inference time using $\mathcal{L}_i$ defined in Equation~\ref{eqn:innerloss} will lead to better healthy tissue segmentation in images with lesions. 
For this purpose we define an \emph{outer loss} to be optimized during meta-learning. 

\begin{figure}[!h]
    \centering
    \includegraphics[width=0.55\textwidth]{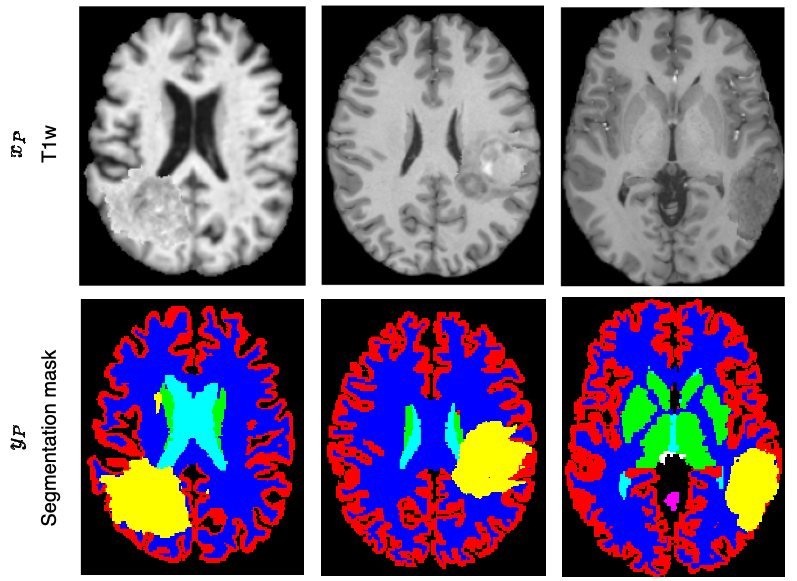}
    \caption{Examples of the pseudo-unhealthy dataset, created by pasting the T1w image voxels with lesion label in BraTS dataset, on top of the lesion free dataset examples; to serve the proxy of test images bearing tumors to adapt to in the meta learning setting.}
    \label{fig:pu}
\end{figure}

Training to achieve better healthy tissue segmentation in images with lesions requires healthy tissue labels in such images, which is exactly what is not available. 
To remedy this, we create a synthetic dataset by artificially introducing lesions on top of images from the lesion-free dataset $X_A$. 
Specifically, we randomly chose a sample from $X_A$, $x_A^{T_1}$ being its T1w image and $y_A$ its ground truth healthy structure segmentation, chose another sample from $D_L$, $x_L^{T_1}$ and $y_L$ being its T1w image and binary tumor segmentation map, respectively. 
We then combine these samples as $x_P^{T_1}\triangleq x_A^{T_1} \odot (1-y_L) + x_L^{T_1} \odot y_L$, where $x_P^{T_1}$ is a \emph{pseudo-lesioned} T1w image.
We also combine the labels $y_P = y_A \odot (1-y_L) + c_L y_L$, where $c_L$ represents the lesion class when combined with the healthy structure classes.
Effectively, the lesion labels replace the ground truth healthy tissue labels in the tumor areas.
The pseudo-lesioned image dataset, where both healthy tissue and lesion labels are available, allows defining an appropriate outer loss. 
Example images and their corresponding labels from the pseudo-unhealthy dataset are shown in Figure \ref{fig:pu}.

In the inner step of the meta learning, we apply the inference time adaptation to pseudo-lesioned images.
We randomly sample two images, one from lesion free $x_A$ and one from pseudo-lesioned datasets $x_P$. 
Then we create an augmented version of the pseudo-lesioned image, i.e., $\tilde{x}$, by randomly placing an integer in the place where lesion voxels are located as described in Section~\ref{sec:problem_formulation}.
Using $\tilde{x}, x_P$ and $x_A$, ideally the inference time optimizes the loss $\mathcal{L}_i$ defined in Equation~\ref{eqn:innerloss}. 
As a full optimization is not possible in the meta-learning approach due to exploding memory requirements, we conclude the inner step by taking one gradient descent step starting from parameters $\theta$ with respect to $\mathcal{L}_i$ and returning parameters $\theta'$.
\begin{equation}
    \theta' = \theta - \alpha \nabla \mathcal{L}_i
\end{equation}

In the outer step, we assess the segmentation performance of $\theta'$ on the pseudo-lesioned image, and update $\theta$ to improve this performance. 
Note that the inner loss maximizes consistency but does not necessarily ensure better segmentation performance. 
Minimization of the following outer loss trains the model to this end.
\begin{equation}
\begin{multlined}
 \mathcal{L}_o = \DSC[g_A(f_{\theta'}(x_P)), (y_A)] 
+\DSC[g_A(f_{\theta'}(\tilde{x})), (y_A)] 
+\DSC[g_A(f_{\theta'}(x_A)), (y_A) ],
\end{multlined}
\end{equation}

Hence, this loss aims to produce accurate anatomical tissue segmentations on all the voxels including those that are pseudo-lesioned. It includes Dice losses on augmented masked images, pseudo-lesioned images and the lesion free images themselves, comparing with the ground truth labels.
The outer loss is maximized with respect to $\theta$, parameters of the T1w feature extractor. 
To this end, at each outer step, $\mathcal{L}_o$ is backpropagated and the model parameters $\theta$ are optimized with Adam optimizer (\cite{adam}).
The only difference between inference time adaptation and the inner steps is that while at inference time, $\mathcal{L}_i$ is minimized with Adam optimizer while we could only use a single gradient step during the inner steps of the meta-learning procedure. 

We would also like to note that this procedure deviates from the original meta-learning approach \cite{maml} described. 
The inner and outer steps described in \cite{maml} uses different samples with the same loss. 
In the procedure described here, the sample is the same but the losses are different. 

\subsubsection{Joint Training}
In the last step of training, we introduce the training of layers $f_\psi$, shown in Figure \ref{fig:model}. 
These layers consist of three convolutions with batch normalization and ReLU \cite{relu} in between. $f_{\psi}$ first generates a spatial attention map $a_F$ using the features from the different branches $w^{T_1}$ and $w^F$. 
To this end, $w^{T_1}$ and $w^F$ are concatenated and passed through convolutional layers to obtain $a_F$. 
Then, to obtain the final segmentation, the features of FLAIR branch is multiplied by the spatial attention and summed with T1 branch features: $w^J = w^{T}+ a_F \odot w^F$. 
Finally, another convolutional layer generates the final segmentation maps $\hat{y}$ following a soft-max function. 

To train the joint model we start with the pretrained $f_\phi$ and the co-trained $f_\theta$, and employ the same meta training scheme for the T1w branch as in the previous step, except, instead of pseudo-lesioned images, we utilize the training set of the BraTS dataset this time, which has T1w and FLAIR sequences available.

In order to be able to use the BraTS dataset in the same meta training scheme, we need healthy tissue segmentations for these images. 
As we mentioned previously, such segmentations do not exist and that is why we used synthetically generated images to get the co-trained $f_{\theta}$ in the previous step. 
Now that we have this $f_{\theta}$, here we use it to generate pseudolabels for the healthy structures for a subset of the BraTS dataset, which we then use as a training set for the joint training step. 
To this end, we simply apply the $f_{\theta}$ with the inference-time adaptation to the T1w images of the BraTS subset. 
To get accurate pseudolabels for the healthy structures, we use the ground truth segmentations for the tumors for the inference adaptation, i.e., we use $y_L$ instead of $\hat{y}_L$ in Equation~\ref{eqn:innerloss}. 
The pseudolabels combined with the manual tumor segmentation form the "ground truth" label for the joint-training.

In the joint training step, parameters of the FLAIR branch $f_\phi$ are kept frozen, and all the other layers are trained.
This step aims to obtain better parameters that allow the features coming from FLAIR branch for the lesion segmentation to be combined with the adapted T1 modality features per image at each iteration.
To this end, first an inner step is taken for the T1w branch's parameters, using the inner loss described in Equation~\ref{eqn:innerloss}, and then the outer loss is optimized with respect to the $\psi$ and $\theta$, effectively using the co-training updates summarized in Algorithm~\ref{alg:meta}.
During the inner loss, we use the manual tumor segmentations, i.e., use $y_L$ instead of $\hat{y}_L$ in Equation~\ref{eqn:innerloss}, as this led to fewer artifacts due to mistakes in the lesion segmentation. We also provide further results of the model that uses $\hat{y}_L$ in the inner step in this training in the Appendix.

\subsection{Joint Inference}
During inference, first the lesion specific sequences of a test sample is passed through the lesion segmentation branch to get an initial prediction of the lesion mask, in our case this is the FLAIR sequence and the mask is obtained by $\hat{y}_L = g_L(f_{\phi}(x^F))$ with trained $g_L$ and $\phi$. 

Then we create the masked augmented image by assigning a random number to the voxels that are segmented as lesion in $\hat{y}_L$.
Inner loss is calculated as described in Equation~\ref{eqn:innerloss} using an additional volunteer image, which does not have any lesions, and it is minimized with the Adam optimizer and parameters of the T1w branch, $\theta$, is updated.
The final prediction is now the output $\hat{y}$ of the model, obtained with spatial attention of both T1 and FLAIR sequence features, i.e., $\hat{y} = f_{\psi}\left(f_{\theta'}(x^{T_1}), f_{\phi}(x^F)\right)$.

Lastly, to increase accuracy and robustness, we follow an ensemble approach. Given a dataset for training, we create 5 random training and validation set splits with random reshuffling. Each model follows all the steps of the training using the different splits, except the pre-training of the T1w branch, yielding 5 different models at the end. Separate inferences are taken from each model. Furthermore, for each model the inference time adaptation is done for 6 different random numbers. This yields in total 30 different predictions for each test sample. The final result of the model is obtained by majority voting through these 30 predictions.

\section{Experiments and Results}
In this section, we explain the details of the experimental setup to assess proposed model's performance and compare with recent alternatives in the literature. 

\begin{table*}[t]
\caption{\label{tab:comparison}Comparison of different methods' Dice scores on the test subjects from BraTS dataset that have been held out as test set. We evaluate the default models of \cite{samseg-lesion} and \cite{billot_synthseg_2021} which are not able to segment gliomas, thus no scores for those models in the tumor row are presented in the joint segmentation case (top). To evaluate healthy tissue segmentation only (down), we copy the ground truth tumor voxels on top of the segmentations output from models. Average Dice over 12 subjects is shown, with standard deviations over 12 subjects in brackets for the joint segmentation table. For healthy tissue segmentation results, average Dice scores and standard deviations of the scores are given over 10 subjects. }
\centering
\scriptsize
\begin{tabular}{l|c|c|c|c|c|c|c|}
\rowcolor{lightgray}\multicolumn{8}{c}{\textbf{Joint segmentation}} \\
\cline{2-8}
 &{\textbf{Gray m.}}  	&	{\textbf{Basal ganglia}} 	&	{\textbf{White m.}}  	&	{\textbf{Tumor}} 	&	{\textbf{Ventricle}} 	&	{\textbf{Cerebellum}} 	&	{\textbf{Brain stem}} \\ \hline
    \multicolumn{1}{|l|}{\textbf{\cite{dorent}}}  	&	 0.840  {(0.046)} &	 0.725 {(0.057)} 	&	 0.883 {(0.020)}  	&	 \textbf{0.899} {(0.038)}  	&	 0.840  {(0.055)} 	&	 0.940  {(0.013)}  	&	 0.775  {(0.025)}  \\ \hline
  \multicolumn{1}{|l|}{\textbf{Ours}} 	&	  \textbf{0.882}  {(0.026)} 	&	 \textbf{0.790}  {(0.046)}	&	 \textbf{0.921}  {(0.017)} 	&	 \textit{0.881}  {(0.064)}  	&	 \textbf{0.879}  {(0.034)}  	&	 \textbf{0.958}  {(0.006)}  	&	 \textbf{0.845}  {(0.041)} \\ \hline
  \multicolumn{8}{c}{}
\end{tabular}
\begin{tabular}{l|c|c|c|c|c|c|c|}
\rowcolor{lightgray}\multicolumn{7}{c}{\textbf{Healthy tissue segmentation}} \\ 
\cline{2-7}
 	&	{\textbf{Gray matter}}  	&	{\textbf{Basal ganglia}} 	&	{\textbf{White matter}}  	&	{\textbf{Ventricle}} 	&	{\textbf{Cerebellum}} 	&	{\textbf{Brain stem}} 	\\ \hline
 \multicolumn{1}{|l|}{\textbf{\cite{dorent}}}  	&	0.849 {(0.045)}	&	0.727 {(0.058)}	&	0.889 {(0.019)}	&	0.838 {(0.061)}	&	0.941 {(0.014)}	&	0.778 {(0.023)}	\\ \hline
 \multicolumn{1}{|l|}{ \textbf{\cite{billot_synthseg_2021}}} 	&	0.819 {(0.024)}	&	0.772 {(0.071)}	&	0.892 {(0.020)}	&	0.858 {(0.044)}	&	0.947 {(0.008)	}&	0.836 {(0.046)}	\\ \hline
 \multicolumn{1}{|l|}{\textbf{\cite{samseg-lesion}}} 	&	0.820 {(0.033)	}&	0.673 {(0.074)	}&	0.880 {(0.022)	}&	0.368 {(0.092)	}&	0.920 {(0.013)	}&	0.797 {(0.040)	}\\ \hline
 \multicolumn{1}{|l|}{\textbf{\cite{kulvbg}}} 	&	0.714 {(0.044)	}&	0.731 {(0.058)	}&	0.863 {(0.027)	}&	0.831 {(0.057)	}&	0.915 { (0.020)	}&	0.765 {(0.079)	}\\ \hline
  \multicolumn{1}{|l|}{\textbf{SynthSR + \cite{billot_synthseg_2021}}} 	&	0.760 {(0.023)	}&	0.747 {(0.063)	}&	0.854 {(0.018)	}&	0.822 {(0.051)	}&	0.935 {(0.009)	}&	0.815 {(0.050)	}\\ \hline
 \multicolumn{1}{|l|}{\textbf{SynthSR + \cite{samseg-lesion}}} 	&	0.722 {(0.036)	}&	0.735 {(0.064)	}&	0.836 {(0.014)	}&	0.799 {(0.063)	}&	0.928 { (0.013)	}&	0.783 {(0.048)	}\\ \hline
 \hline
\multicolumn{1}{|l|}{\textbf{Ours}} 	&	\textbf{0.889} {(0.029)	}&	\textbf{0.792} {(0.052)	}&	\textbf{0.925} {(0.020)	}&	\textbf{0.878} {(0.040)	}&	\textbf{0.959} {(0.007)	}&	\textbf{0.844} {(0.048)	}\\ \hline
\end{tabular}
\end{table*}

\begin{table*}[t]
\caption{\label{tab:comparison-ours}Comparison of the different steps of our proposed models' Dice scores on the test subjects from BraTS dataset that have been held out as test set. Average Dice over 12 subjects is shown, with standard deviations over 12 subjects in brackets for the joint segmentation table. For healthy tissue segmentation results, average Dice scores and standard deviations of the scores are given over 10 subjects. }
\centering
\scriptsize
\begin{tabular}{l|c|c|c|c|c|c|c|}
\rowcolor{lightgray}\multicolumn{8}{c}{\textbf{Joint segmentation}} \\
\cline{2-8}
 &{\textbf{Gray m.}}  	&	{\textbf{Basal ganglia}} 	&	{\textbf{White m.}}  	&	{\textbf{Tumor}} 	&	{\textbf{Ventricle}} 	&	{\textbf{Cerebellum}} 	&	{\textbf{Brain stem}} \\ \hline   
  \multicolumn{1}{|l|}{\textbf{Pretrained}} 	&	 {0.873} {(0.021)} 	&	\textbf{0.792}  {(0.054)}  	&	 {0.918} {(0.017)} 	&	 \textbf{0.884}  {(0.042)} 	&	 {0.875}  {(0.039)} 	&	 0.950  {(0.017)} 	&	 0.839  {(0.048)}  \\ \hline
  \multicolumn{1}{|l|}{\textbf{Ours Co-tr.}}&	 0.873  {(0.033)}	&	 0.789  {(0.053)}	&	0.913  {(0.021)}  	&	 \textbf{0.884}  {(0.042)} 	&	 0.873  {(0.035)}  	&	 0.958  {(0.007)}  	&	 \textbf{0.845} {(0.041)}  \\ \hline
  \multicolumn{1}{|l|}{\textbf{Ours Joint tr.}} 	&	  \textbf{0.882}  {(0.026)} 	&	 {0.790}  {(0.046)}	&	 \textbf{0.921} {(0.017)} 	&	 \textit{0.881}  {(0.064)}  	&	 \textbf{0.879}  {(0.034)}  	&	 \textbf{0.958}  {(0.006)}  	&	 \textbf{0.845} {(0.041)} \\ \hline
  \multicolumn{8}{c}{}
\end{tabular}
\begin{tabular}{l|c|c|c|c|c|c|c|}
\rowcolor{lightgray}\multicolumn{7}{c}{\textbf{Healthy tissue segmentation}} \\ 
\cline{2-7}
 	&	{\textbf{Gray matter}}  	&	{\textbf{Basal ganglia}} 	&	{\textbf{White matter}}  	&	{\textbf{Ventricle}} 	&	{\textbf{Cerebellum}} 	&	{\textbf{Brain stem}} 	\\ \hline
 \multicolumn{1}{|l|}{\textbf{Pretrained}} 	&	0.880 {(0.020)}	&	\textbf{0.797} {(0.054)}	&	0.924 {(0.017)}&	0.878 {(0.042)}	&	0.954 {(0.011)}&	0.839 {(0.051)}\\ \hline
 \multicolumn{1}{|l|}{\textbf{Ours Co-trained}} 	&	0.883 {(0.032)}	&	0.795 {(0.049)}&	0.920 {(0.021)}	&	0.875 {(0.036)}	&	\textbf{0.960} {(0.007)}	&	0.844 {(0.044)}	\\ \hline
\multicolumn{1}{|l|}{\textbf{Ours Joint trained}} 	&	\textbf{0.889} {(0.029)	}&	{0.792} {(0.052)	}&	\textbf{0.925} {(0.020)	}&	\textbf{0.878} {(0.040)	}&	{0.959} {(0.007)	}&	\textbf{0.844} {(0.048)	}\\ \hline
\end{tabular}
\end{table*}

\begin{figure}[!ht]
    \centering 
    \includegraphics[width=0.95\textwidth]{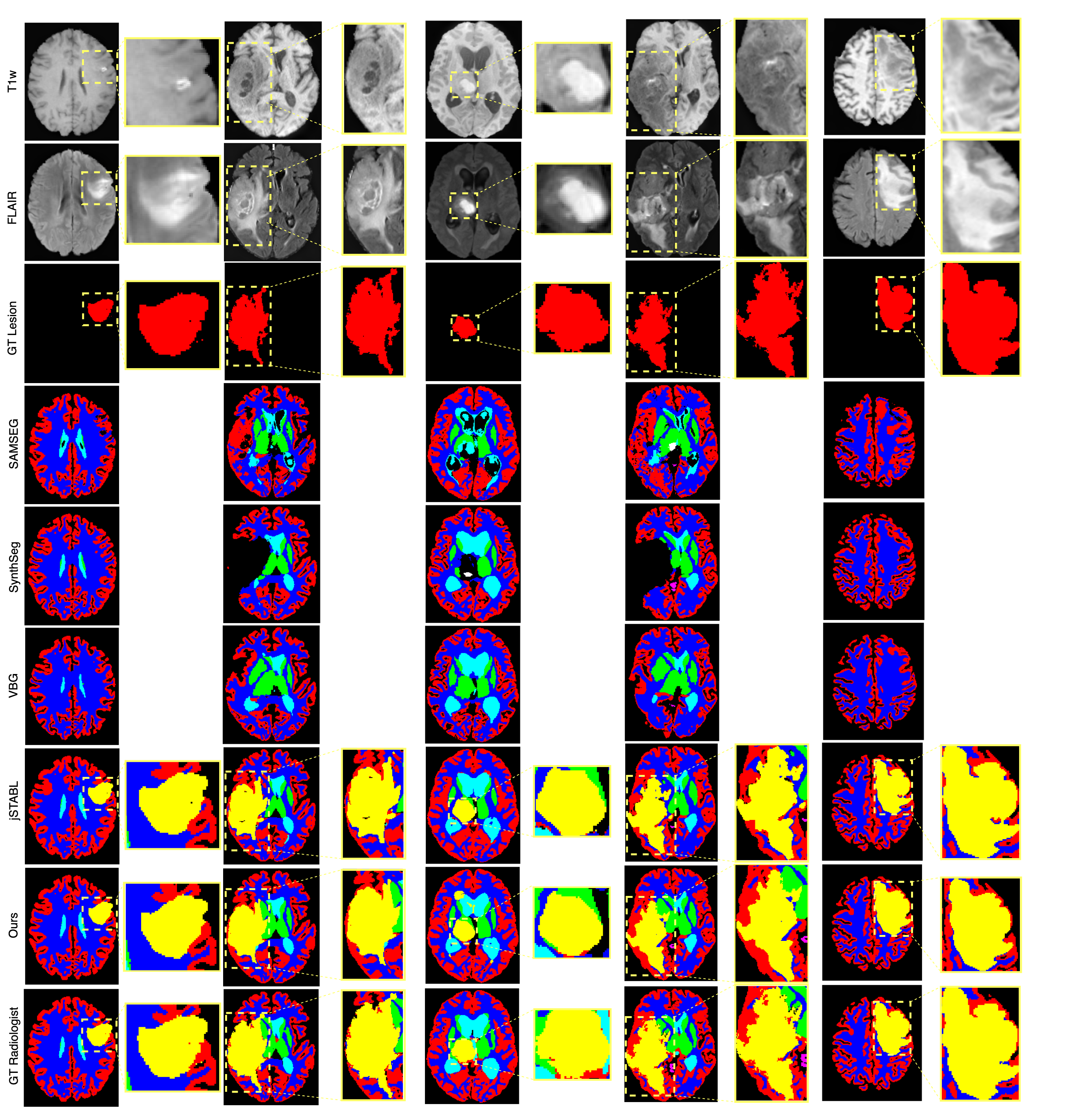}
    \caption{Comparison of different methods on slices selected from 5 test subjects. First three rows are given in the BraTS dataset, T1w image highlights tissues, whereas in FLAIR lesions are more visible. Tumor subclasses are merged into one class as shown in row 3. Next four rows show competing methods for healthy structure segmentation. We show our model's segmentation in penultimate column, and the last column shows the lineation of classes by radiologist, which is taken as the ground truth for the quantitative analysis. SAMSEG \cite{samseg-lesion}, SynthSeg \cite{billot_synthseg_2021} and VBG \cite{kulvbg} do not produce tumor segmentation results.}
    \label{fig:comparison}
\end{figure}

\begin{figure}
    \centering
    \includegraphics[width=0.8\textwidth]{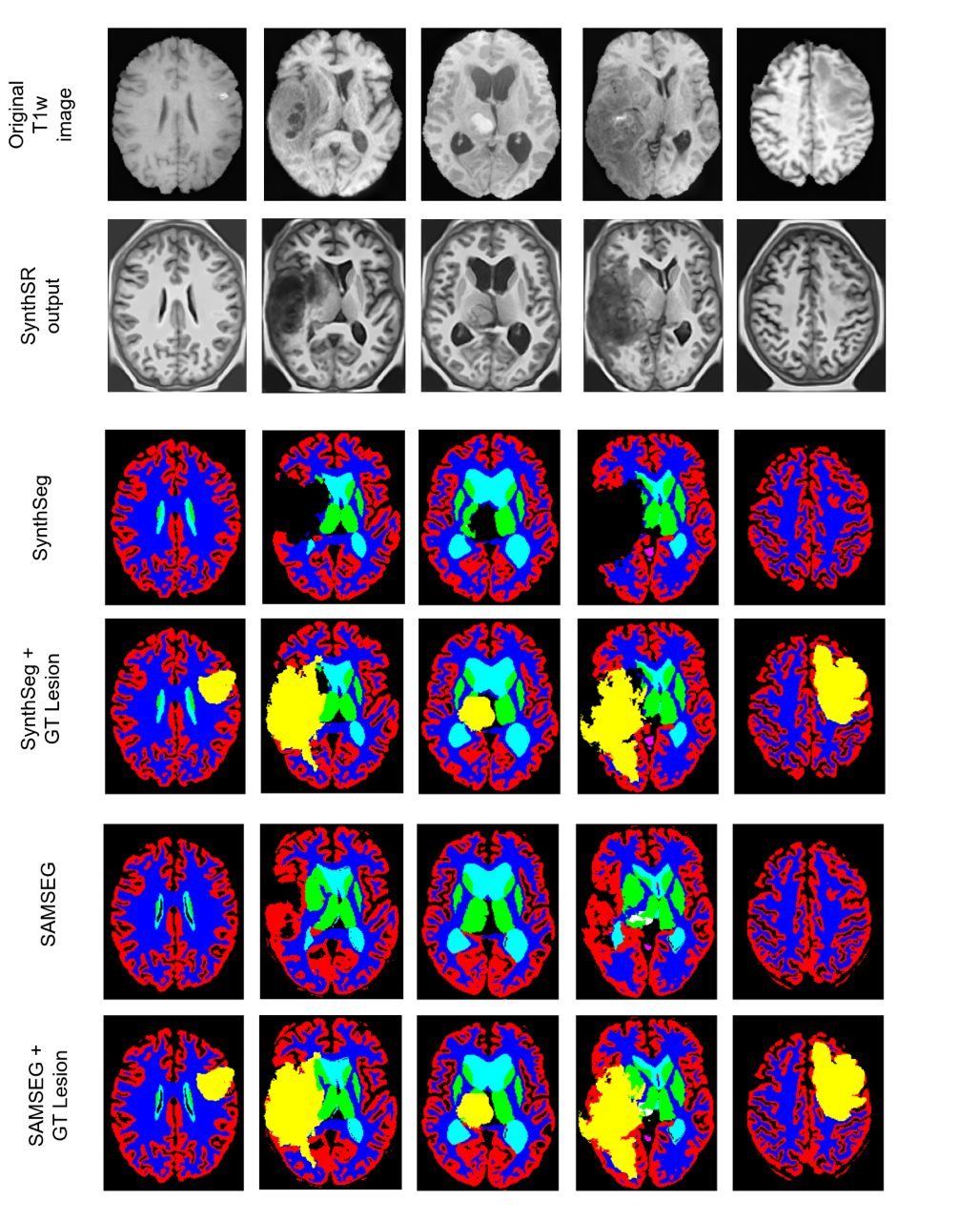}
    \caption{Qualitative results for SynthSeg and SAMSEG after the original T1w image has transformed with SynthSR. The first row shows the original T1w images from the test image set. The second row shows the output of SynthSR method, where the small lesions should be inpainted. SynthSeg output, and the ground truth lesion annotation overlaid, SAMSEG output and the ground truth lesion annotation overlaid for the same images are shown respectively in the remaining rows. }
    \label{fig:synthsr}
\end{figure}

\subsection{Datasets}
We utilize three datasets for our experiments. 
Two datasets for lesion free subjects $\mathbf{X}_A$, Human Connectome Project (HCP) \cite{hcp} and Alzheimer's Disease Neuroimaging Initiative (ADNI) \cite{mueller2005alzheimer} datasets; and one for subjects with lesions $\mathbf{X}_L$, Multimodal Brain Tumor Image Segmentation Benchmark (BraTS) 2018 \cite{brats1, brats2, brats3},  dataset.
 The ADNI was launched in 2003 as a public-private partnership, led by Principal Investigator Michael W. Weiner, MD. The original goal of ADNI was to test whether serial magnetic resonance imaging (MRI), positron emission tomography (PET), other biological markers, and clinical and neuropsychological assessment can be combined to measure the progression of mild cognitive impairment (MCI) and early Alzheimer's disease (AD). The current goals include validating biomarkers for clinical trials, improving the generalizability of ADNI data by increasing diversity in the participant cohort, and to provide data concerning the diagnosis and progression of Alzheimer’s disease to the scientific community. For up-to-date information, see adni.loni.usc.edu.

We use the T1 weighted (T1w) images from HCP and ADNI datasets to serve the purpose of lesion-free subject images and T1w and FLAIR images of BraTS 2018 dataset for patient images with lesions.
The reason to include the ADNI dataset along with the HCP is to have more generalization capability over elderly subjects.
Adding these subjects increases the variety of appearances of atrophies and enlarged ventricles due to aging and neurodegenerative processes.
We selected 22 subjects from the elderly subjects of ADNI dataset and 30 subjects from HCP dataset to curate the lesion-free training dataset.
We obtain ``ground truth'' segmentations of healthy structures for these selected images using Freesurfer \cite{freesurfer}. 
While these are not necessarily ground truth as they have not been generated by an algorithm and not an expert, reliability and accuracy of Freesurfer segmentations have been established in the literature for high quality T1w images for HCP and ADNI datasets, and we visually check the segmentations to make sure they are of good quality and have no big errors. 
Finally, the labels created using Freesurfer were reduced down to 7 classes for demonstration purposes: gray matter, white matter, basal ganglia, ventricle, cerebellum, brain stem and background. 

114 subjects were selected from BraTS 2018 dataset. 
12 of them were kept for hold-out testing and the remaining have been split to training and validation sets of 90 and 12 subjects. 
The split has been repeated 5 times after random reshuffling.
For the 102 subjects used in training and validation splits, only the lesion segmentation were available. 
Healthy structure segmentations were not available. 
The labels for different parts of the tumor available in the BraTS dataset (necrotic core, edema, enhancing core, non enhancing core) were reduced down to one label of lesion.
For the 12 samples kept for hold-out testing, lesion segmentations were available and, additionally, we obtained ground-truth segmentations of the healthy structures for quantitative evaluation. 
To obtain these ground truth segmentation, first an initial segmentation of the healthy structures were obtained with a 2D U-Net \cite{unet} model trained with lesion free images to segment the structures targeted in this work. The output of this model on the test images with lesion are obtained, then the initial segmentations were manually corrected. 
This correction step was performed by a professional segmentation and labeling company (Labelata GmbH, Zürich, Switzerland) whose network includes radiologists with sub-speciality experience in neuro-radiology.
The algorithm that generated the initial segmentations was a 2D U-Net \cite{unet} trained with images from the HCP \cite{hcp} dataset were used. 

Additionally, we utilized an in house brain MRI patient dataset from Städtische Klinikum Dessau, Germany to demonstrate generalization performance. 
The dataset contains MRI images of 48 patients with glioblastoma WHO IV. 
All patients underwent primary resection and postoperative radiotherapy.
The dataset of each of the 48 patients consisted of one MRI scan series before resection, all postoperative imaging series, and the radiation therapy treatment protocol.
The availability of modalities vary among the patients in the dataset, we utilize the T1w and FLAIR sequences for this work.

\subsubsection{Preprocessing}
We employed affine registration to MNI space \cite{mni} using ANTsPy \cite{ants}, skull stripping using HD-BET software \cite{isensee2019automated}, bias field correction \cite{tustison2010n4itk} and cropping to the smallest volume containing the brain and z-score normalization as preprocessing steps before inputting images to the network.
We used the same preprocessing to compare with other models and the quantitative results were calculated on the preprocessed version of the test subjects.

\subsection{Training Details}
The three stages of training, namely pretraining with task specific datasets, meta co-training and joint training are done for $150$, $100$, and $50$ epochs respectively. 
We used Adam optimizer for pretraining and optimized outer losses in both meta and joint training parts with a learning rate of $0.001$.
For the inner loss during meta-learning, we used gradient descent following previous literature with a step size of $0.005$ \cite{maml}. 
We used a batch size of $2$, due to high memory constraints of the meta learning scheme combined with 3D operations on images.
Due to memory constraints, we operate on 3D patches of size  $128^3$   instead of the whole images.
For testing, we utilize the same patch size with 20 voxels overlap in between patches.
Effectively, each inner and outer step operates on one patch. 
Meta-updates are performed only when lesion is present in the patch.
Random numbers used in masking augmentation are selected from the set $\{-5,-2,-1,1,2,5\}$.

U-Net style networks used in $f_\theta$ and $f_\phi$ have 5 contracting and upsampling blocks with filters starting from $16$ and doubling up to $256$. 
We employ random brightness and contrast augmentations in training.
Dice losses were calculated including the batch dimension (reducing in all dimensions except class dimension) and then averaged.
Training took in total $\sim 42$ hours on one NVIDIA A100 GPU.
Testing takes $\sim 5$ minutes per subject volume.
We used pytorch \cite{pytorch} for deep learning libraries, and torchmeta \cite{torchmeta} for meta learnable layers.
We used majority voting for the $5$ folds used in cross validation in the pretraining of lesion model only.
Testing the meta learned models includes: running each model with a random value for the masked augmentation, then majority voting on all random values and all $5$ folds to obtain final predictions.

\subsection{Results} 
We show and compare the performance of our proposed model on the 12 hold-out test samples. 
We additionally evaluated the model on an in house hold-out dataset.
\subsubsection{Comparisons}
We compared the proposed method with four alternative methods on the 12 hold-out test samples.

\textbf{Sequence Adaptive Multimodal SEGmentation (SAMSEG)} \cite{samseg-lesion} segments brain structures from multi-contrast MRI data.
SAMSEG algorithm relies on a forward probabilistic model, and inverting them to obtain the segmentation via a Bayesian approach.
For one scan, the inference takes about 15 minutes.
The authors propose an additional model that can jointly segment multiple sclerosis lesions as well, however; our preliminary evaluations with that model showed that the model does not generalize for disruptive lesions such as brain tumors.
Therefore we use the variation of the model without the capability of segmenting multiple sclerosis lesions.
On the other hand, the original version worked well for the healthy structures in T1w images bearing tumors. 
So, we opted for using the original version that was made available in the Freesurfer software suite in our evaluations. 
SAMSEG model segmented the healthy structures but not the lesion nor the area under the lesion.
No preprocessing is needed for SAMSEG and the models are made available publicly.

\textbf{SynthSeg} \cite{billot_synthseg_2021} is a brain structure segmentation model for any contrast and resolution.
They use domain randomization, by generating a wide range of synthetic data from segmentations and train a 3D convolutional neural network.
Similar to SAMSEG, SynthSeg has an additional model for white matter hyperintensities that is not completely fitting with this type of lesions, as evaluations on our test set showed.
However, as the SAMSEG model, the original SynthSeg model performed well on segmenting healthy structures in T1w images bearing tumors. 
So, we also opted for using the original version of the algorithm in our evaluations, also for segmenting the healthy structures outside the lesion area. 
Both SAMSEG and SynthSeg predicts more labels than we used. 
We employed the same merging strategy used while creating the training set to obtain 6 healthy tissue classes as predictions. 

For the models SynthSeg and SAMSEG, we make and additional evaluation with another model with lesion inpainting.
We employ the model proposed by \cite{synthsr}, called SynthSR, to obtain 1 mm isotropic MP-RAGE scans for the test volumes. We then run SynthSeg and SAMSEG on these volumes to get segmentations.
Authors in \cite{kulvbg} proposed \textbf{Virtual Brain Grafting (VBG)} as a reliable method for segmenting brain MR images with a broad spectrum of brain lesions. 
The method takes T1w image and the lesion mask as inputs, and goes through a pipeline of creating a pseudohealthy T1w image with lesion filling, then using Freesurfer's recon-all pipeline to generate segmentations.
The pipeline expects the ground truth segmentation of the lesions to create the generated lesion free brain, which is different than the other methods.
We used the model from the official repository to generate segmentations for the 12 hold-out test images from BraTS dataset.
The software failed on 2 of these images, therefore, we omit these failures and report the average Dice scores for the healthy tissue segmentation for the remaining 10 subjects in Table \ref{tab:comparison}.

Finally, \textbf{jSTABL}, proposed in \cite{dorent}, aims to tackle the joint segmentation problem using task specific hetero modal datasets.
They provide a variational formulation for the joint problem, and propose different approaches for images bearing gliomas and images with multiple sclerosis lesions.
For images bearing glioma, the model trained jointly with control (healthy) scans and images with lesion perform well and they obtain even better results with introducing domain adaptation methods and creating pseudo healthy versions of the glioma images.
We use their pretrained model available in their repository for glioma images, which expects T1w, FLAIR, T1w with contrast and T2w images together as input.
The pretrained model's training dataset is not exactly corresponding to our train / test splits, which gives slightly unfair advantage to their method.
While using this tool, we provide all the sequences to jSTABL. 

\subsubsection{Results on BraTS dataset}
We show the Dice scores over 7 classes in Tables \ref{tab:comparison} and \ref{tab:comparison-ours} for the 12 images in the test set.
The Dice scores are calculated with respect to the radiologist segmentations.
Table on the top shows results on joint segmentation task of lesions and healthy tissues, whereas the bottom table shows results on tissue segmentation only for both tables. 
The latter results are based on Dice calculations that assumes voxels remaining in the lesion are correctly segmented. 
We show the evolution of our models' Dice scores through its different stages in the Table \ref{tab:comparison-ours}.
The row denoted as ``pretrained'' refers to the combination of task specific models pretrained with their respective datasets.
To obtain the joint segmentation, lesion segmentations are predicted as the majority vote of 5 pre-trained lesion segmentation models, i.e., pre-trained $g_L \circ f_\phi$ models trained with random training and validation splits using $\mathbf{X}_L$. 
These predictions are then are pasted on segmentations coming from the pre-trained T1w network, i.e. pre-trained $g_A \circ f_\theta$ trained on $\mathbf{X}_A$. 
The combination of these two task specific pre-trained models already create a strong baseline for the next stages. 
Yet, referring to problem demonstrated in Figure \ref{fig:healthyseg}, we have unreliable segmentations around lesions.
Since the segmentation of healthy tissue structures from the T1w branch are unreliable, putting the lesion segmentation coming from the FLAIR branch $f_\phi$ affects the final tissue segmentation drastically.

\begin{figure*}[]
    \centering
    \includegraphics[width=0.7\textwidth]{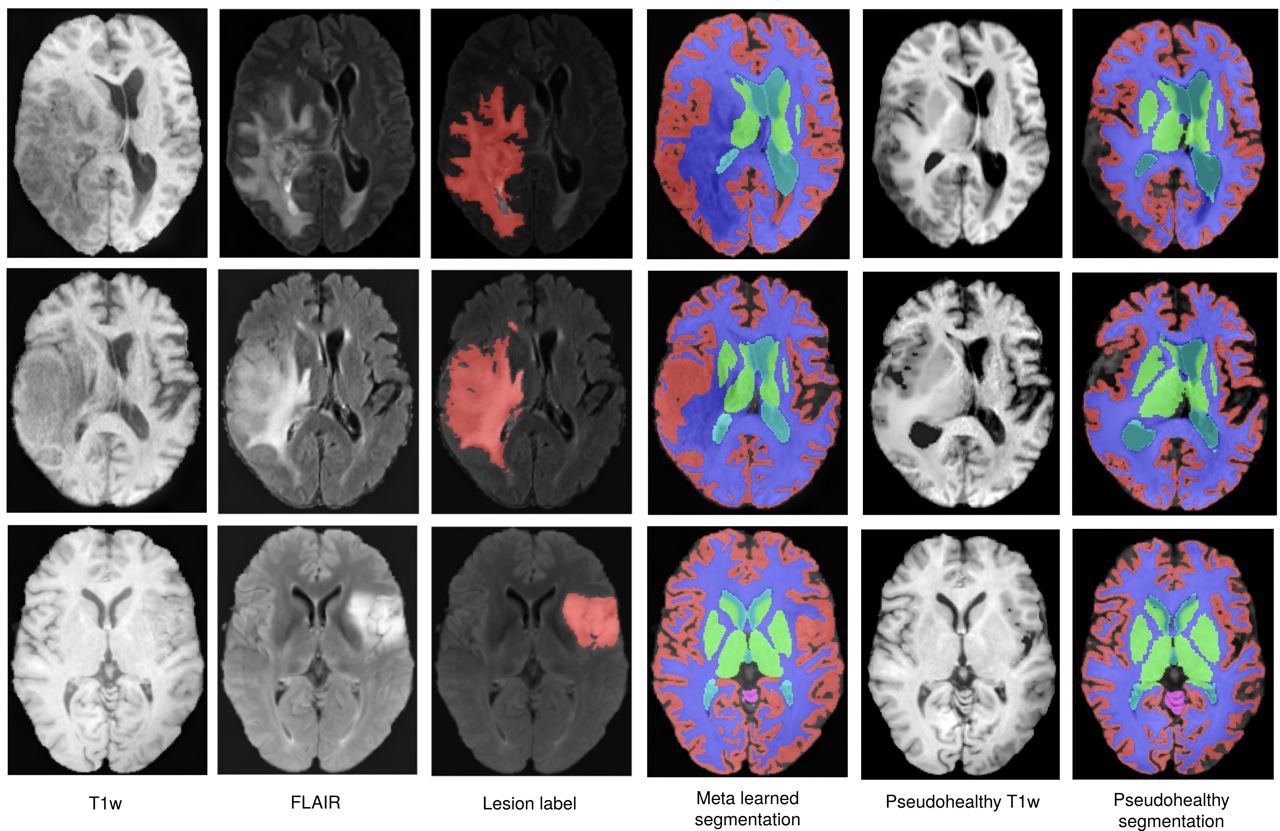}
    \caption{Examples of images from the test set and healthy tissue segmentation performance of our proposed model and \cite{kulvbg}'s virtual brain grafting method. First three columns correspond to original images and the lesion mask from BraTS dataset, fourth column is the healthy segmentation overlaid obtained after meta-training step in our proposed pipeline. The fifth column correspond to the grafted brain by the method, with the overlaid freesurfer segmentation output on the right.}
    \label{fig:vbg}
\end{figure*}

Next rows in Table \ref{tab:comparison-ours} shows the model performance after the second stage: meta co-training.
This step aims to obtain healthy tissue segmentations by enforcing model to ignore the regions within lesions, and not fail in the presence of big and disruptive lesions.
Dice scores of this stage show similar performance in almost all classes compared to the pre-trained model.
This performance is not surprising, since the meta co-trained model has only seen images with synthetically added lesions during its training. 
When applied to images with real tumors, we expect the model to encounter difficulties due to possible domain shifts. 
However, while this intermediate step demonstrates not much better performance compared to the pre-trained model, it is a crucial step for preparation for the joint model training. It allows the model to (i) generate pseudohealthy versions of the tumor images, which is then used for training the joint model, and (ii) be adaptable to the presence of lesions for the joint segmentation stage.
Results in Figures~\ref{fig:vbg} and~\ref{fig:clinical} highlight this point and will be discussed shortly. 

Finally, the Dice scores of our proposed model after joint training is shown in the last rows of Tables \ref{tab:comparison} and \ref{tab:comparison-ours}.
The model outperforms the compared methods in most of the classes, performing second only in the lesion segmentation. 
\cite{dorent} shows better performance in this class by a small margin, which could be due to using all available scans from the BraTS dataset (T1, T1c, T2 and FLAIR), whereas our model uses only FLAIR modality for the lesion segmentation. 

Visual results on five test images can be seen in Figure \ref{fig:comparison}. 
Here, in the fourth row,  we see that \cite{samseg-lesion} was able to delineate the white-gray matter border but failed to adapt to large ventricle sizes and their intensity ranges in unhealthy images.
On row 5 of Figure \ref{fig:comparison}, we show \cite{billot_synthseg_2021}, which uses domain randomization in training.
This model was successful in segmenting healthy structures, as demonstrated in the images. 
The lesion parts were missing or segmented incorrectly, which, in any case, is not expected from this model. 
However, even outside the lesion area, the model was still influenced by the adverse affects of the disruptive lesion as demonstrated by the quantitative results in Table \ref{tab:comparison}. 
This may be due to discrepancies especially in areas close to the lesion, where the area containing the lesion is segmented as the background rather than any other tissue class. 
We show the quantitative  results of \cite{samseg-lesion} and \cite{billot_synthseg_2021} in Table \ref{tab:comparison} after applying the \cite{synthsr} method for lesion inpainting. 
The Dice scores for the healthy segmentation seem to decrease for \cite{billot_synthseg_2021} in most classes and increase for \cite{samseg-lesion} using this inpainting technique before segmentation. 
Especially for the ventricle class, the output from SynthSR creates as better input for the method of \cite{samseg-lesion}.
The overall segmentation and the segmentation around lesions however, are still subpar to the other methods with the MP-RAGE generation.
First main reason is the increase of contrast between the gray matter and white matter after the image generation.
This change of contrast results in an undersegmented gray matter in general for both SAMSEG and SynthSeg, and results in a lower dice score for these classes, since the ground truth segmentations are obtained using the original T1w images.
We show the qualitative results in \ref{fig:synthsr} for the same images as in Figure \ref{fig:comparison}. 
As seen in the Figure \ref{fig:synthsr}, the second reason why the segmentations are not as successful is the process of lesion inpainting. 
For the bigger lesions such as in column 4, the inpainting could not remove the anomalous parts, and further increased the contrast and changed the lesion border, resulting in a bigger discrepancy with the ground truth segmentations, as seen in row 4.
Overall, we can say that using SynthSR before segmentation was useful, especially for SAMSEG, but the original image and the lesion segmentation should also be taken into account when this method is applied.

Results of \cite{kulvbg} are shown in the sixth row, which are also quite good. 
The model was able to generate healthy tissue segmentations even in areas covered by the lesion, thanks to generation of the pseudohealthy image. 
However, the virtual brain grafting failed where the lesions were very big and the generated image became smaller as the part of the brain was reconstructed as background (on column 4, for example).
We show some example outputs from the pipeline of \cite{kulvbg} in Figure \ref{fig:vbg}, illustrating the filled T1w images with respect to given lesion masks.
Here, we compare the output of the \cite{kulvbg} with out proposed model's T1w branch output after meta co-training stage. 
The presence of a less disruptive lesion as shown in row 3, results in a better grafting output than the first two rows and the segmentation is more reliable. 
However, examples with grossly disruptive and irregular lesions, such as those shown in the first two rows, are harder to reliably fill and the voxels on the background gray matter border become generally undersegmented.
Our model manages to delineate these borders better and create more proportionally appropriate structures compared to the rest of the image, as seen on the larger cyan (ventricle) parts compared to VBG segmentation output.

Finally, we compare visually with the work closest to ours, \cite{dorent} in row seven of Figure \ref{fig:comparison}. 
We can interpret the Dice scores in Table \ref{tab:comparison} better by looking at these images.
Generally, \cite{dorent} tended to oversegment the gray matter compared to our method, which resulted in lack of details, visible in all columns.
Our model tended to oversegment lesions, which resulted in a lower Dice score in tumor class. 
Furthermore, \cite{dorent}'s segmentation of brain stem and white matter border had a different border consensus than what we used in training as seen in the second example, which lowered the performance on that class significantly compared to the radiologist segmentations. 
Differences in these structures should be taken with a grain of salt. 
Our model showed greater performance in areas closer to bigger lesions, which could be seen in the fourth column, around the right and top of the lesion, where the ground truth is basal ganglia and \cite{dorent}'s output was white matter.
We also note that ground truth radiologist segmentations were corrections on an automated segmentation obtained from U-Net, and not performed by multiple experts, this could introduce biases and imperfections.

\subsubsection{Results on the in-house dataset} \label{inhouse}
We further demonstrate a possible use case of our method on a different private longitudinal dataset.
This in house dataset consists of multiple sequences at multiple time points for patients admitted to hospital with brain tumors.
Segmentation of healthy tissues accurately might help with positioning the brain and assessing the changes.
Thus, the proposed model could potentially be useful, if we can get consistent segmentations for the same subject over different time points.

The dataset provided to us by the hospital has 5 to 10 time points for each patient with varying number of different sequences.
We utilized our joint model trained with the BraTS, HCP and ADNI datasets. 
Since the dataset is completely unlabeled, we utilized the pretrained FLAIR branch of our model with parameters $f_\phi$ and $g_L$ to obtain lesion segmentations.
Some of these samples showed different intensity characteristics compared to our training data for lesion segmentation. 
We observed hypointense parts in the lesion, which had not been seen by our segmentation model, and lower contrast overall between the lesion and the brain structure in the FLAIR sequences.
These differences led to worse lesion segmentations predictions than what we observed on the hold-out test samples, which originate from the BraTS dataset. 

\begin{figure}[h!]
    \centering
    \includegraphics[width=0.7\textwidth]{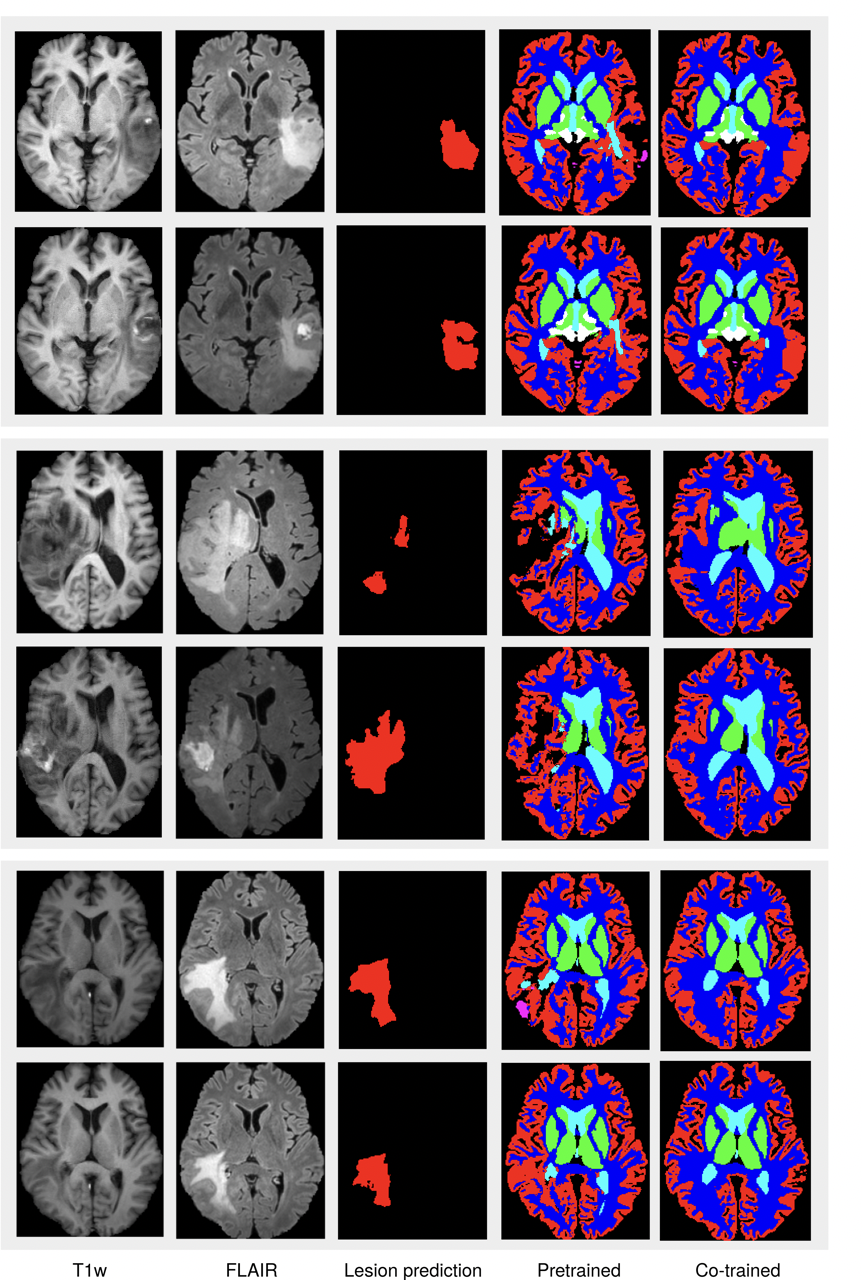}
    \caption{Examples of three subjects from the in house dataset with two scans each in different timepoints. First two columns show the T1w and FLAIR sequences from the dataset. The images on the top rows for each subject are taken before operation and the bottom row images are taken after an operation. Third column shows the lesion segmentation obtained from FLAIR branch of our proposed model, trained with BraTS images. Penultimate column shows the output of pretrained model with volunteer images, and the last column shows the meta co-trained model, with pseudo unhealthy images that have the lesions of BraTS training dataset. We show the consistency of the proposed model's segmentations on the last column, despite the domain gap between the training and testing datasets, and imperfect lesion segmentations. }
    \label{fig:clinical}
\end{figure}

We show some visual examples in Figure \ref{fig:clinical} for three different subjects. 
For each subject, we show pre-operation and post-operation images taken 4 to 7 days  apart. 
Column 3 shows the lesion segmentations obtained from the FLAIR sequences, and the adverse effects of out of domain appearance of lesions and the contrast changes are visible on the segmented images, specifically for the second subject (middle). 
Segmentation outputs of the models in the pretrained and meta co-trained stages are shown in the last two columns. 
From these segmentations, we can say that the meta co-trained model produces consistent segmentations over these two time points. 
Moreover, consistency is higher compared to results of the pre-trained model. 
Importantly, we observed a decrease in the quality of segmentation when the lesion was not segmented correctly, in row 3 of group 2 for examples where basal ganglia (green) is missing.

\begin{figure}
    \centering
    \includegraphics[width=0.8\linewidth]{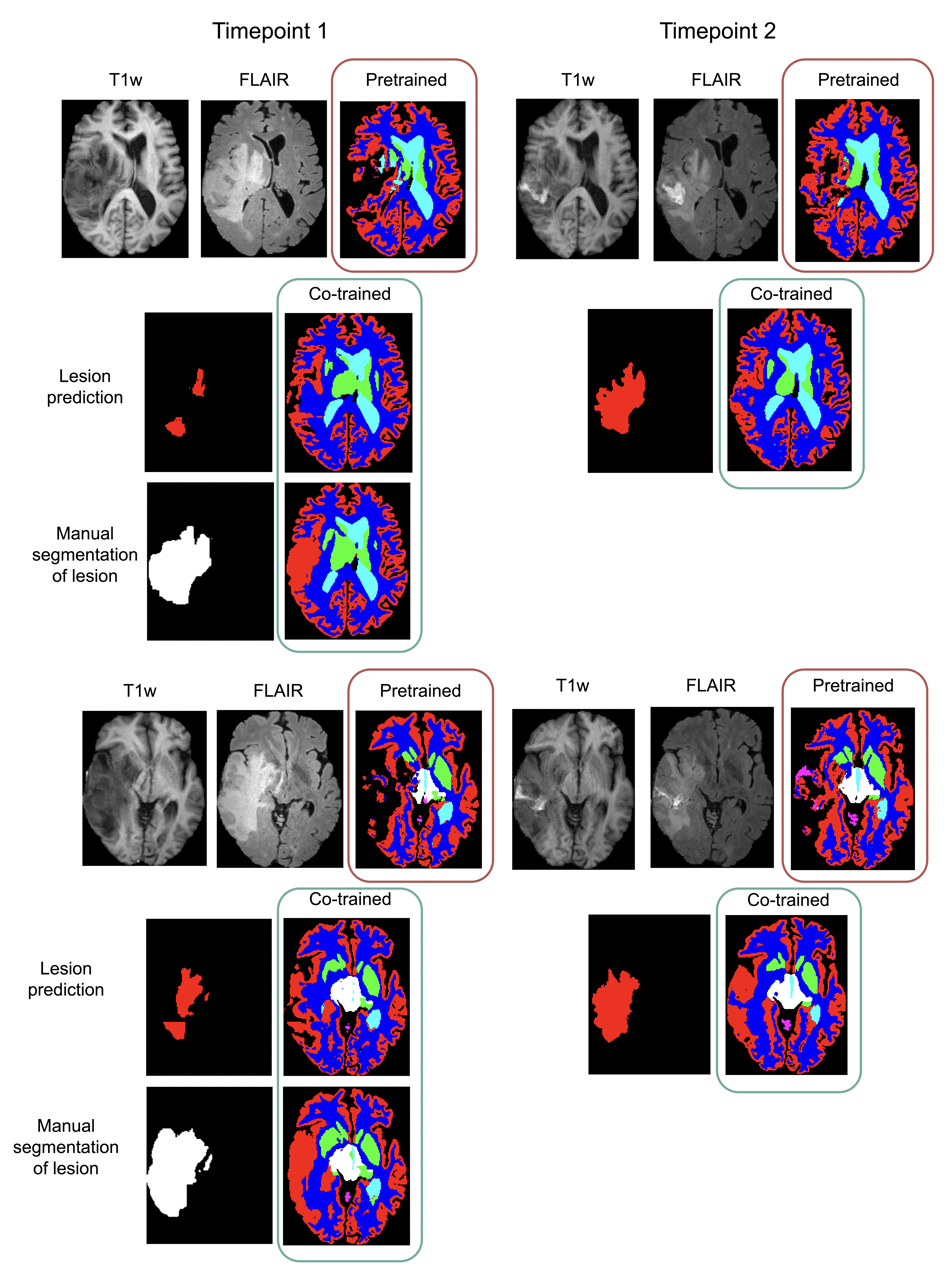}
    \caption{Segmentation results for one subject from the in house dataset for two slices. First row shows the T1w and FLAIR images and the segmentation obtained with the pretrained model. Red lesion segmentations are the output of our lesion segmentation model trained with BraTS dataset, and the segmentation to the right show the co-trained model output with respect to using these as lesion masks. Last row with white color segmentations are done manually and the segmentations with these are shown to the right. }
    \label{fig:clinical-manual}
\end{figure}

Overall, the proposed model performed well on the in house samples, which showed domain differences, in particular acquisition shifts. 
We believe that using manually drawn accurate lesion segmentations, the predictions for the healthy tissue structures would improve even further. 
Here we utilized the model trained with BraTS dataset, however, for future work adding augmentations to create similar appearances for the lesions on these dataset could be useful for more accurate lesion segmentations and hence better anatomical tissue segmentations.
To demonstrate the possible effect of a more accurate lesion segmentation on the co-trained meta model, we manually segment the lesion on the second subject shown in Figure \ref{fig:clinical}.
In Figure \ref{fig:clinical-manual}, we show that using a better lesion segmentation mask for the testing of co-trained model can improve the healthy segmentation performance. Specifically, the white-gray matter border and the basal ganglia segmentations improve and become more robust, therefore a better correspondence can be obtained within different timepoints.

\section{Discussion}

\begin{figure}[!h]
    \centering
    \includegraphics[width=0.65\textwidth]{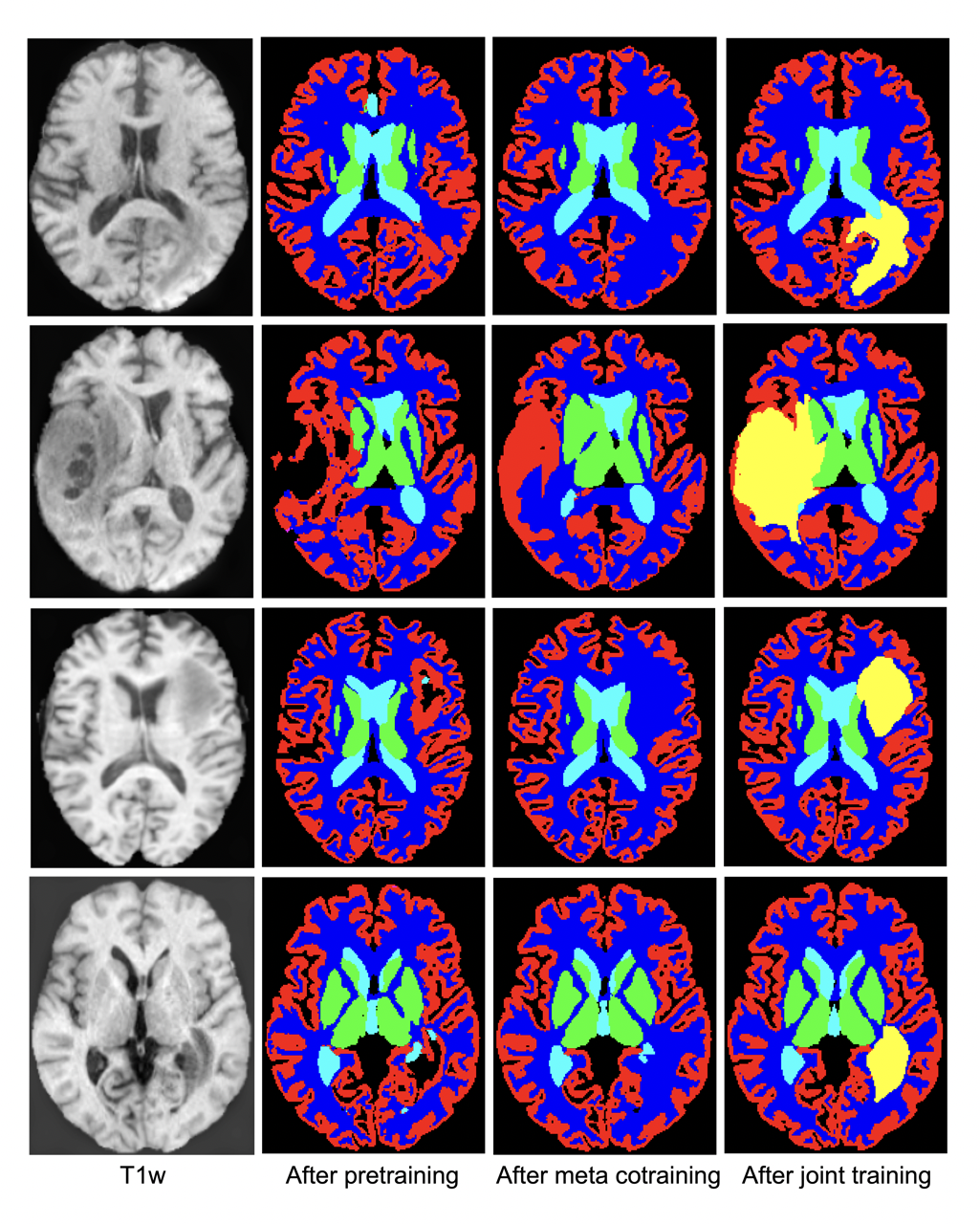}
    \caption{Intermediate results of the model on BraTS test images, first column denotes the T1 weighted image input, second, third and fourth columns show the output segmentation of the pretrained model, meta co-trained model and the joint model respectively. }
    \label{fig:prevsmeta}
\end{figure}

In Table \ref{tab:comparison-ours}, we demonstrated the performance difference between the different training stages in our proposed model: pretrained, meta co-trained and jointly trained model.
Qualitative comparison of the test image predictions between pretraining, meta co-training and joint training stages are shown in Figure \ref{fig:prevsmeta}. 
Healthy tissue segmentation is placed where lesions are existing after meta training, whereas pretrained model generally has confusion with background in predictions for these voxels.

\begin{figure}[!ht]
    \centering
    \includegraphics[width=0.45\textwidth]{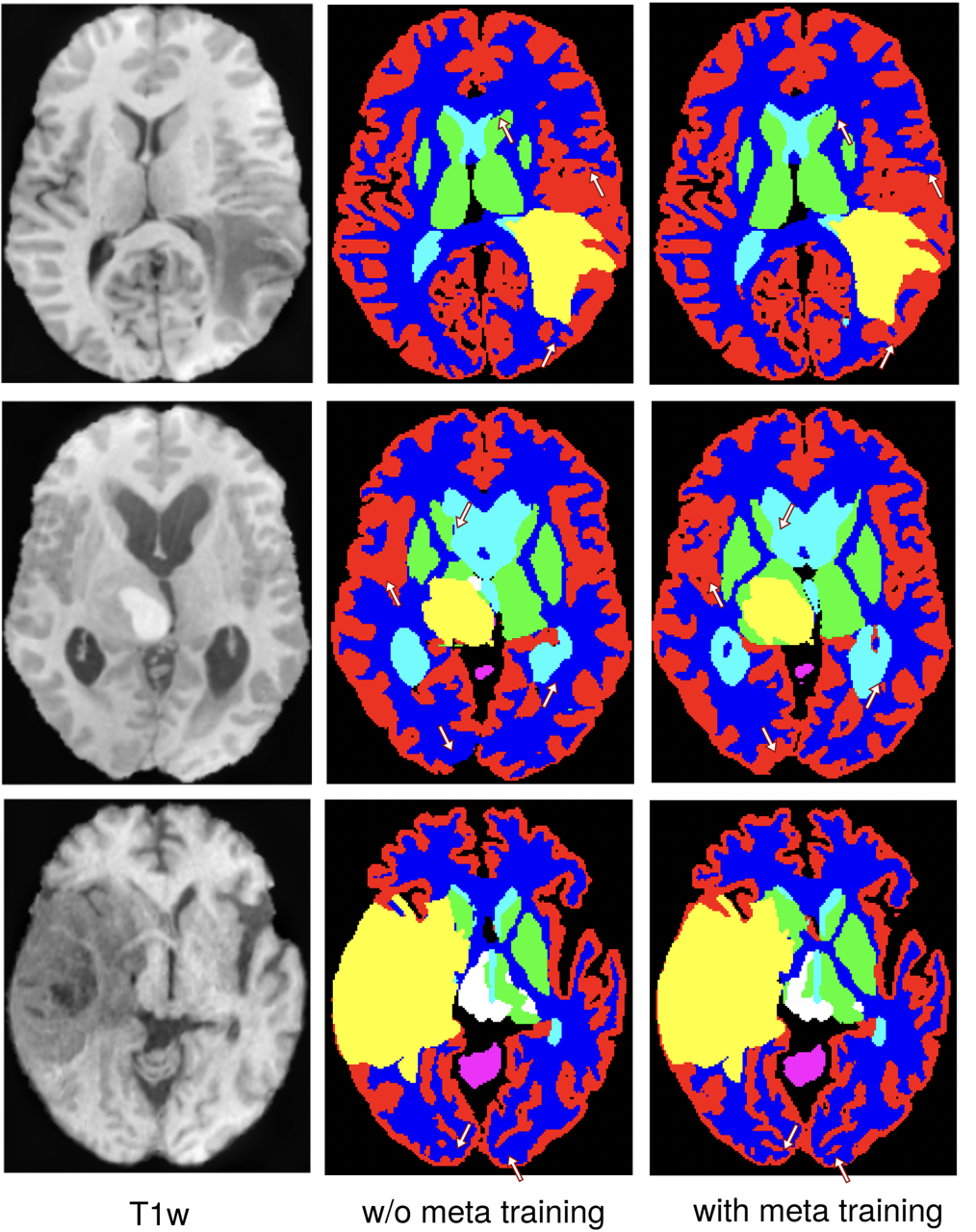}
    \caption{Comparison of segmentation results of BraTS test images showing not including the inner loop ($L_i$) vs training with $L_i$ in meta learning fashion. Example in first row shows a less drastic intensity change in an around lesion, whereas the examples in last two rows show more irregular lesions. Correspondingly, healthy tissue segmentation around lesion does not change as much in the first row compared to the last two rows, whereas improvement of detail of the gray/white matter border with the meta training is visible in all examples. }
    \label{fig:adapt}
\end{figure}

We show the significance of this meta learning scheme in Figure \ref{fig:adapt}.
This visualization is obtained by taking the outputs of $g_A\circ f_\theta$ for both models, and merging the output of $g_L\circ f_\phi$ pretrained with lesion dataset $x_L$ for FLAIR modality inputs (thus they are the same in both models).
If meta training is omitted in this step, i.e., we minimize only the outer loss $\mathcal{L}_o $ with respect to all the model parameters $\theta$, we get better segmentations than the pre-trained models.
This is because the outer loss directly maximizes the Dice score between model predictions for pseudo-unhealthy images and the corresponding healthy tissue labels, including areas within the lesion. 
Yet, significant details are still missing in this optimization without the adaptation, as seen in all rows in Figure \ref{fig:adapt}. 
The border between gray and white matter is lacking details, the enlarged ventricles (which do not exist as frequently as they do in test set compared to healthy training set) are undersegmented (shown in color cyan).

\begin{figure*}[h!]
    \centering
    \includegraphics[width=\textwidth]{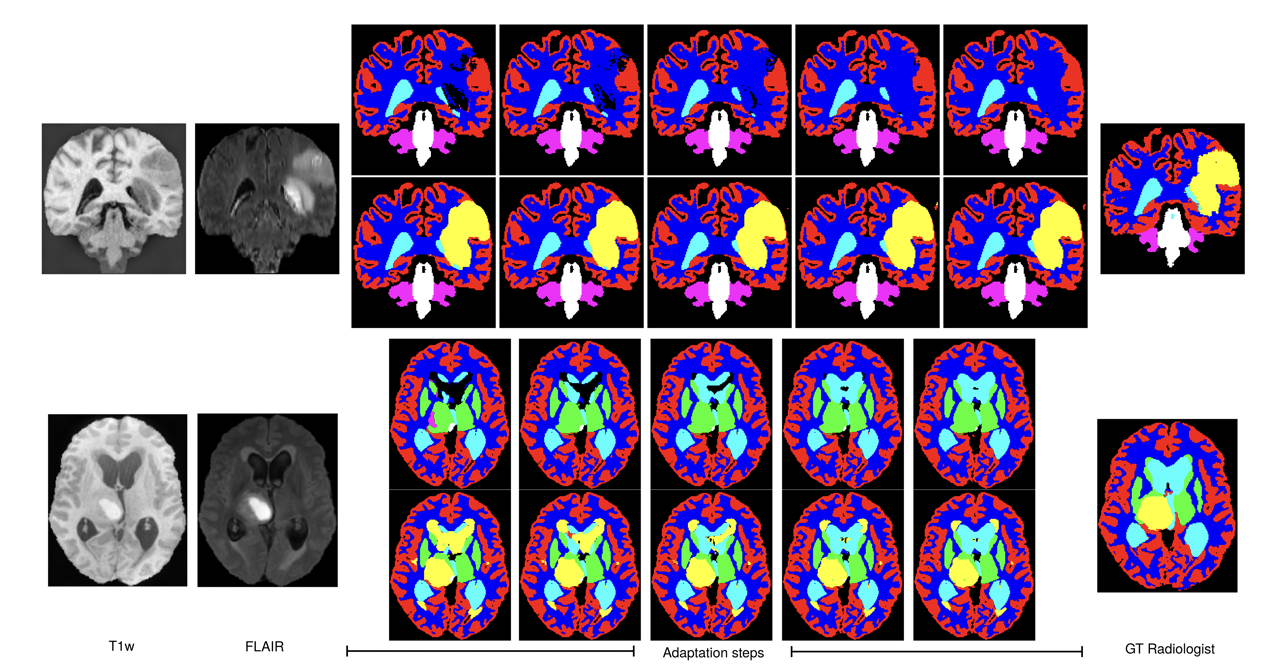}
    \caption{Progression of segmentations through adaptation steps of the joint inference for two subjects from the test set, top row shows the anatomical segmentation branch output, bottom row shows joint segmentation output. The subject at the top of the figure shows less disruption on anatomical segmentation due to presence of the lesion compared to the subject at the bottom. }
    \label{fig:metasteps}
\end{figure*}

Meta co-trained and joint model results were obtained after adapting to each test sample, and the adaptation steps to get to the final segmentations are depicted in Figure \ref{fig:metasteps}.
The segmentations on the first rows for each image show the anatomical segmentation branch output ($g_A\circ f(x^{T_1},\theta_j)$) and the lower rows show the joint segmentation output.
We observe that segmentations that are initially reliable do not change as much as the optimization steps advance, since the consistency between the original image and the mask-augmented version is already high.
In test cases where the consistency is lower, such as the last row in Figure \ref{fig:metasteps}, segmentation are corrected through adaptation steps to get to a better point in the parameter space for that test sample.
Specifically, in the adaptation steps, we maximize the anatomical segmentation consistency with respect to augmentations on the lesion.
After the joint training with this consistency, we observe that the anatomical segmentation on the lesion voxels are adapted to a point where the fused features and the joint segmentation performance would be higher than the starting point.

We have showed the qualitative results for an in house dataset in section \ref{inhouse}.
For these patients it is of interest how the brain structures change before and after operations, and as the brain adapts and heals during recovery.
Inspecting and analyzing the changes through time points require the registered and matched view for the expert, which is challenging due to drastic changes in appearance of the brain around tumor and resection area.
The segmentation results showed consistency over different time points for the out of distribution samples. 
This suggest that the proposed model has good generalization capability out of the box, but can be further improved by including examples from variety of appearances of different lesions.
Since the model has separate feature extractors for the different sequences (i.e. T1w and FLAIR), it could also be easily adapted to include more sequence types, which might improve the generalization capability of the lesion segmentation part, allowing the tissue segmentation obtained from the common sequence to be more reliable.
Doing so would require additional encoders to output features in a similar fashion as the ones used here. 

Our experiments and results on the BraTS dataset showed promising performance on segmentation of tissues and lesions jointly for images with gliomas.
The  healthy tissue segmentation performance was assessed on the seven labels annotated by radiologists, yet atlas-based methods such as \cite{samseg-lesion} and \cite{kulvbg} can produce sub cortical level segmentations.
The registration pipeline to obtain segmentations for these models is very time consuming.
Contrarily, convolutional neural network based models such as \cite{billot_synthseg_2021}, \cite{dorent} and our method  have the advantage of faster inference (hours vs seconds).
Since our model runs an optimization in the test time, the time consumed in the inference may result in the order of minutes, with respect to the number of steps in the optimization.
Furthermore, optimizing our proposed model requires a significant amount of memory due to calculating second order gradients for the meta learnable parameters.
We have employed a training scheme with limited samples per optimization step, which could be improved.

In this work, we utilized the availability of different sequences per subject for both volunteer and patient datasets and required these two datasets to share one sequence where target healthy tissue structures can be segmented.
Differently in \cite{dorent}, the authors utilized all available sequences for the unhealthy dataset, while allowing for possibility of missing sequences through information averaging. 
Incorporating all sequences by averaging results in similar or worse performance as seen by the qualitative results in Figure \ref{fig:comparison}
and Table \ref{tab:comparison}. 

\section{Conclusions}

In this work, we tackled the problem of anatomical tissue segmentation in images bearing disruptive lesions. 
We utilized task specific and disparately labeled datasets, and viewed the problem in two separate parts: anatomical segmentation and lesion segmentation, each having their respective task specific network architectures.
First, we used images with lesion to train lesion segmentation network, and lesion free images to train anatomical segmentation network.
Then, we proposed a meta learning strategy to adapt the anatomical segmentation network so that it could segment healthy structures even in the presence of disruptive lesions. 
To this end, we generated images with fake lesions to simulate test time adaptation of the anatomical segmentation model to minimize adverse affects of the lesion area on the healthy structure predictions. 
We optimized the anatomical segmentation network to produce outputs based on the performance of this test time task.
Finally, we combined the task specific network features using a spatial attention mechanism.
We showed results on a public dataset with glioblastoma patients and an in house dataset.
While improving joint segmentation performance for lesions and healthy tissue structures, we observed that this training strategy could also be used to generate pseudo-healthy versions of original images bearing tumors. 

\section{Author Contributions}
Meva Himmetoglu: Conceptualization, formal analysis, investigation, methodology, software, validation, visualization, writing and editing.

Ilja Ciernik: Data curation. 

Ender Konukoglu: Conceptualization, funding acquisition, supervision, methodology, writing - review and editing.

\section{Acknowledgements}
This project is funded by the Swiss National Science Foundation (SNF) (205320\_200877).
We thank Yessin Rafael Boudersa for efforts in clinical data curation.
We acknowledge and thank Emiljo Mëhillaj and Georg Brunner for project discussions and assistance.

Data collection and sharing for the Alzheimer's Disease Neuroimaging Initiative (ADNI) is funded by the National Institute on Aging (National Institutes of Health Grant U19AG024904). The grantee organization is the Northern California Institute for Research and Education. In the past, ADNI has also received funding from the National Institute of Biomedical Imaging and Bioengineering, the Canadian Institutes of Health Research, and private sector contributions through the Foundation for the National Institutes of Health (FNIH) including generous contributions from the following: AbbVie, Alzheimer’s Association; Alzheimer’s Drug Discovery Foundation; Araclon Biotech; BioClinica, Inc.; Biogen; Bristol- Myers Squibb Company; CereSpir, Inc.; Cogstate; Eisai Inc.; Elan Pharmaceuticals, Inc.; Eli Lilly and Company; EuroImmun; F. Hoffmann-La Roche Ltd and its affiliated company Genentech, Inc.; Fujirebio; GE Healthcare; IXICO Ltd.; Janssen Alzheimer Immunotherapy Research \& Development, LLC.; Johnson \& Johnson Pharmaceutical Research \& Development LLC.; Lumosity; Lundbeck; Merck \& Co., Inc.; Meso Scale Diagnostics, LLC.; NeuroRx Research; Neurotrack Technologies; Novartis Pharmaceuticals Corporation; Pfizer Inc.; Piramal Imaging; Servier; Takeda Pharmaceutical Company; and Transition Therapeutics.

\bibliographystyle{plain} 
\bibliography{bibliography}

\appendix
\section{Labeling Convention} \label{sec:labeling}
We utilize Freesurfer labels for training examples from lesion free datasets.
The Freesurfer labels can be found in the following link: \url{https://surfer.nmr.mgh.harvard.edu/fswiki/FsTutorial/AnatomicalROI/FreeSurferColorLUT}
We use the following mapping function to reduce the number of labels:
\begin{lstlisting}[language=Python, caption=Python function to assign labels for Freesurfer labels]
def assign_labels(label_data):
    """
    Free surfer Label numbers : https://surfer.nmr.mgh.harvard.edu/fswiki/FsTutorial/AnatomicalROI/FreeSurferColorLUT

    0: 'background',
    # 1: 'cortical gray matter',
    # 2: 'basal ganglia',
    # 3: 'white matter',
    # 4: 'lesion',
    # 5: 'ventricles',
    # 6: 'cerebellum',
    # 7: 'brain stem'
    """

    relabelled_data = label_data.copy()
    unique_element_ids = np.unique(relabelled_data)

    for ele_id in unique_element_ids:
        if ele_id in [0, 1, 24, 6, 40, 45, 15]:
            # background
            relabelled_data[label_data == ele_id] = 0
        elif ele_id in [2, 41, 251, 252, 253, 254, 255, 30, 62, 77]:
            # white matter
            relabelled_data[label_data == ele_id] = 3
        elif ele_id in [9, 10, 11, 12, 13, 17, 18, 26, 48, 49, 50, 51, 52, 53, 54, 58]:
            # basal ganglia
            relabelled_data[label_data == ele_id] = 2
        elif ele_id in [25, 57]:
            # lesion
            relabelled_data[label_data == ele_id] = 4
        elif ele_id in [4, 5, 14, 43, 44, 72, 31, 63]:
            # ventricles
            relabelled_data[label_data == ele_id] = 5
        elif ele_id in [7, 8, 46, 47]:
            # cerebellum
            relabelled_data[label_data == ele_id] = 6
        elif ele_id in [16, 28, 60]:
            # brain stem
            relabelled_data[label_data == ele_id] = 7
        else:
            # cortical gray matter
            relabelled_data[label_data == ele_id] = 1

    return relabelled_data
\end{lstlisting}
\newpage
\section{Dataset generation} \label{sec:pseudounhealthy}
\begin{figure}[h!]
    \centering
    \includegraphics[width=0.6\linewidth]{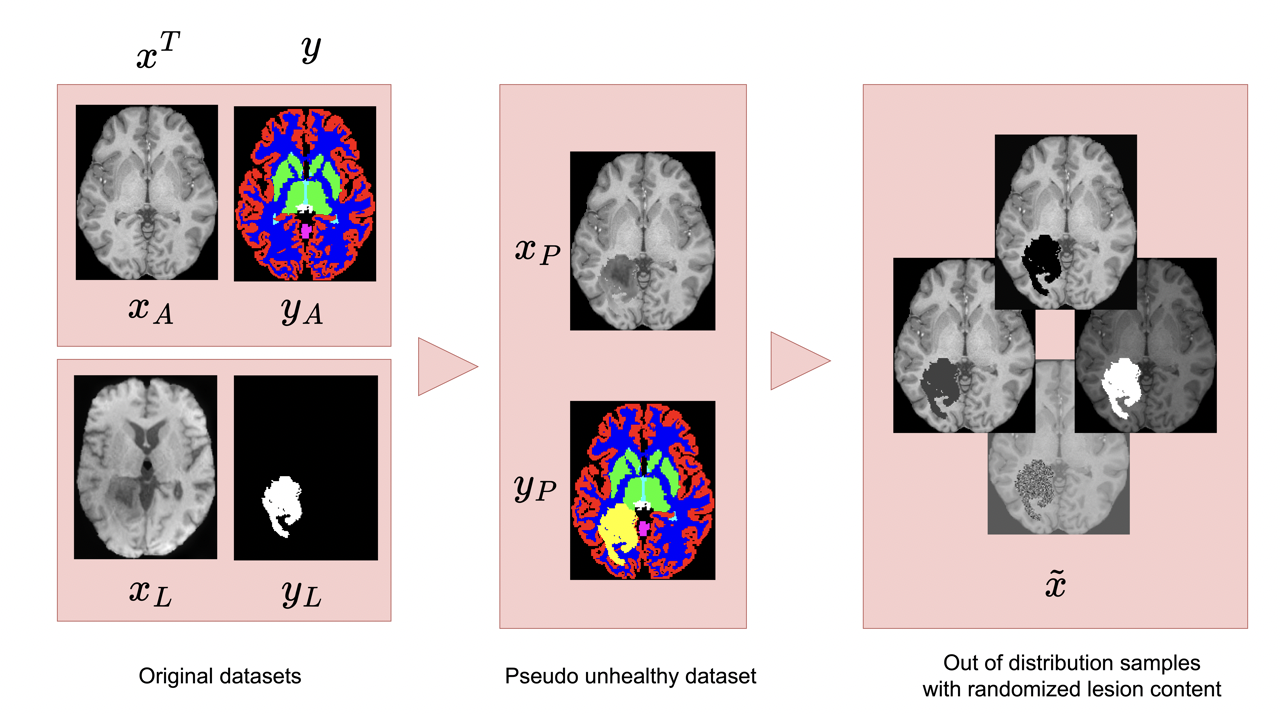}
    \caption{This figure explains the dataset generation used in meta co-training phase. We start with dataset with lesions $x_L$ and lesion free dataset $x_A$, which are co-registered. We take the pixels from T1w images with lesion that are labeled as lesion, and replace them in the lesion free dataset images. We do the same in the label space to create $x_P$ and $y_P$ pairs of pseudo unhealthy dataset. Finally, to create out of distribution samples $\tilde{x}$, we set the \textit{fake} lesion pixels to different values. }
    \label{fig:pseudounhealthy-generation}
\end{figure}

\section{Segmentation Results in Hausdorff Distance Metric}

\begin{figure}[h]
    \centering
    \includegraphics[width=0.7\linewidth]{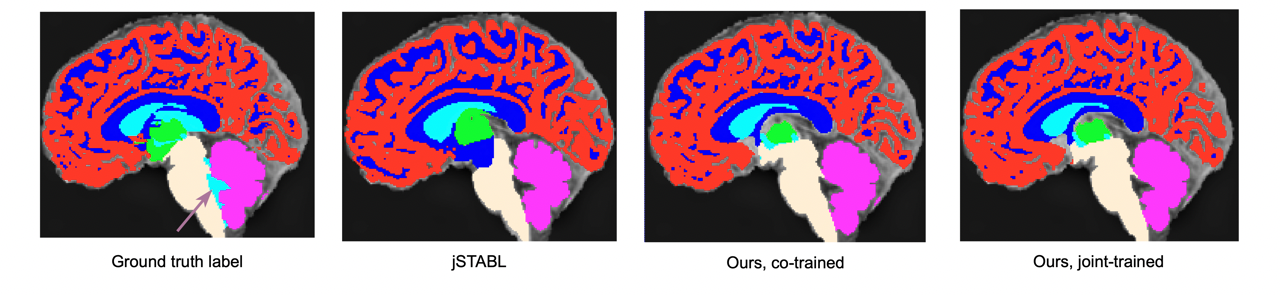}
    \caption{Example segmentation results over sagittal plane. The ventricle annotation for the 4th ventricle does not exist in our training set, thus our conversion from Freesurfer labels to target classes, but exists in the radiologist labels which we use as ground truth; resulting in drastically large numbers for Hausdorff distances for all models.}
    \label{fig:ventricle}
\end{figure}

We share the results comparison of different methods on the test subjects of BraTS dataset, with the Hausdorff distance metric at 95 percentile.
We note that for ventricle class, the reported values are quite high, and we show the reasoning visually in Figure \ref{fig:ventricle}. In the conversion method from Freesurfer labels to target classes, we choose to put 4th ventricle in the background class, yet the ground truth we obtain from radiologists include this structure under the label ventricle. This mismatch in labeling causes all models to have bigger than expected Hausdorff distance for the class ventricle.

\begin{table*}[h!]
\caption{\label{tab:comparison-hd}Comparison of different methods' Hausdorff distances at 95 percentile on the test subjects from BraTS dataset that have been held out as test set. Average Dice over 12 subjects is shown, with standard deviations over 12 subjects in brackets for the joint segmentation table. For healthy tissue segmentation results, average Dice scores and standard deviations of the scores are given over 10 subjects. }
\centering
\scriptsize
\begin{tabular}{l|c|c|c|c|c|c|c|}
\rowcolor{lightgray}\multicolumn{8}{c}{\textbf{Joint segmentation}} \\
\cline{2-8}
 &{\textbf{Gray m.}}  	&	{\textbf{Basal ganglia}} 	&	{\textbf{White m.}}  	&	{\textbf{Tumor}} 	&	{\textbf{Ventricle}} 	&	{\textbf{Cerebellum}} 	&	{\textbf{Brain stem}} \\ \hline
    \multicolumn{1}{|l|}{\textbf{\cite{dorent}}}  	&	 1.75 {(0.31)} &	 7.11  {(2.32)} 	&	 2.13 {(0.37)}  	&	 \textbf{8.15} {(10.90)}  	&	 40.14  {(7.21)} 	&	 4.34  {(1.17)}  	&	 12.01  {(3.22)}  \\ \hline
    \hline
  \multicolumn{1}{|l|}{\textbf{Pretrained}} 	&	 {1.72} {(0.49)} 	&	\textbf{5.96}  {(2.10)}  	&	 {\textbf{1.63}}  {(0.34)} 	&	23.59  {(32.59)} 	&	 \textbf{34.96}  {(9.47)} 	&	 15.37  {(22.50)} 	&	 14.83  {(16.15)}  \\ \hline
  \multicolumn{1}{|l|}{\textbf{Ours Co-tr.}}&	1.63 {(0.32)} & 6.22 {(1.81)} & 1.85 {(0.51)} & {23.59} {(32.59)} & 37.81 {(9.29)} & 3.60 {(0.96)} & 7.96 {(2.90)} \\ \hline
  \multicolumn{1}{|l|}{\textbf{Ours Joint tr.}} 	& \textbf{1.59} {(0.36)} & 6.29 {(2.19)} & 1.74 {(0.48)} & {11.64} {(13.15)} & 37.34 {(9.04)} & \textbf{3.17} {(0.93)} & \textbf{7.95} {(2.77)} \\ \hline
  \multicolumn{8}{c}{}
\end{tabular}
\begin{tabular}{l|c|c|c|c|c|c|c|}
\rowcolor{lightgray}\multicolumn{7}{c}{\textbf{Healthy tissue segmentation}} \\ 
\cline{2-7}
 	&	{\textbf{Gray matter}}  	&	{\textbf{Basal ganglia}} 	&	{\textbf{White matter}}  	&	{\textbf{Ventricle}} 	&	{\textbf{Cerebellum}} 	&	{\textbf{Brain stem}} 	\\ \hline
 \multicolumn{1}{|l|}{\textbf{\cite{dorent}}}  & 1.70 {(0.35)} & 7.20 {(2.46)} & 2.01 {(0.43)} & 40.23 {(7.87)} & 4.53 {(1.20)} & 12.27 {(3.48)} \\ \hline
 \multicolumn{1}{|l|}{ \textbf{\cite{billot_synthseg_2021}}} 	& 2.25 {(1.57)} & 7.06 {(3.01)} & 2.18 {(0.71)}  & 37.55 {(8.49)} & 4.78 {(0.66)} & 8.32 {(3.44)} \\ \hline
 \multicolumn{1}{|l|}{\textbf{\cite{samseg-lesion}}} 	& 2.00 {(0.47)} & 6.14 {(2.00)} & 2.31 {(0.43)}  & 38.00 {(3.59)} & 5.67 {(1.58)} & \textbf{7.48} {(3.87)} \\ \hline
 \multicolumn{1}{|l|}{\textbf{\cite{kulvbg}}} & 3.15 {(1.54)} & 6.49 {(2.29)} & 2.54 {(1.03)}  & 39.3 {(3.84)} & 4.17 {(1.48)} & 8.05 {(4.60)} \\ \hline
  \multicolumn{1}{|l|}{\textbf{SynthSR + \cite{billot_synthseg_2021}}} 	& 2.37 {(1.10)} & 7.03 {(2.19)} & 2.54 {(0.53)}  & 38.00 {(1.76)} & 5.06 {(1.76)} & 8.82 {(3.31)} \\ \hline
 \multicolumn{1}{|l|}{\textbf{SynthSR + \cite{samseg-lesion}}} & 2.46 {(0.83)} & 6.36 {(2.21)} & 3.45 {(0.58)}  & \textbf{35.82} {(9.60)} & 5.10 {(1.14)} & 8.65 {(3.75)} \\ \hline
 \hline
 \multicolumn{1}{|l|}{\textbf{Pretrained}} 	& 1.75 {(0.49)} & \textbf{6.04} {(2.33)} & \textbf{1.55} {(0.35)} &  \textbf{35.82} {(9.92)} & 16.72 {(24.01)} & 7.83 {(2.28)} \\ \hline
 \multicolumn{1}{|l|}{\textbf{Ours Co-trained}} 	& \textbf{1.50} {(0.43)} & 6.22 {(2.10)} & 1.63 {(0.53)} & 37.14 {(37.14)} & 3.24 {(3.24)} & 8.06 {(3.04)} \\ \hline
\multicolumn{1}{|l|}{\textbf{Ours Joint trained}} & 1.57 {(0.37)} & 6.34 {(2.38)} & 1.67 {(0.47)}  & 37.03 {(9.78)} & \textbf{3.22} {(1.01)} & 8.08 {(3.00)} \\ \hline
\end{tabular}
\end{table*}

\newpage
\section{Pseudolabel Generation for Joint Training}
\begin{figure} [h!]
\centering
    \includegraphics[width=0.85\linewidth]{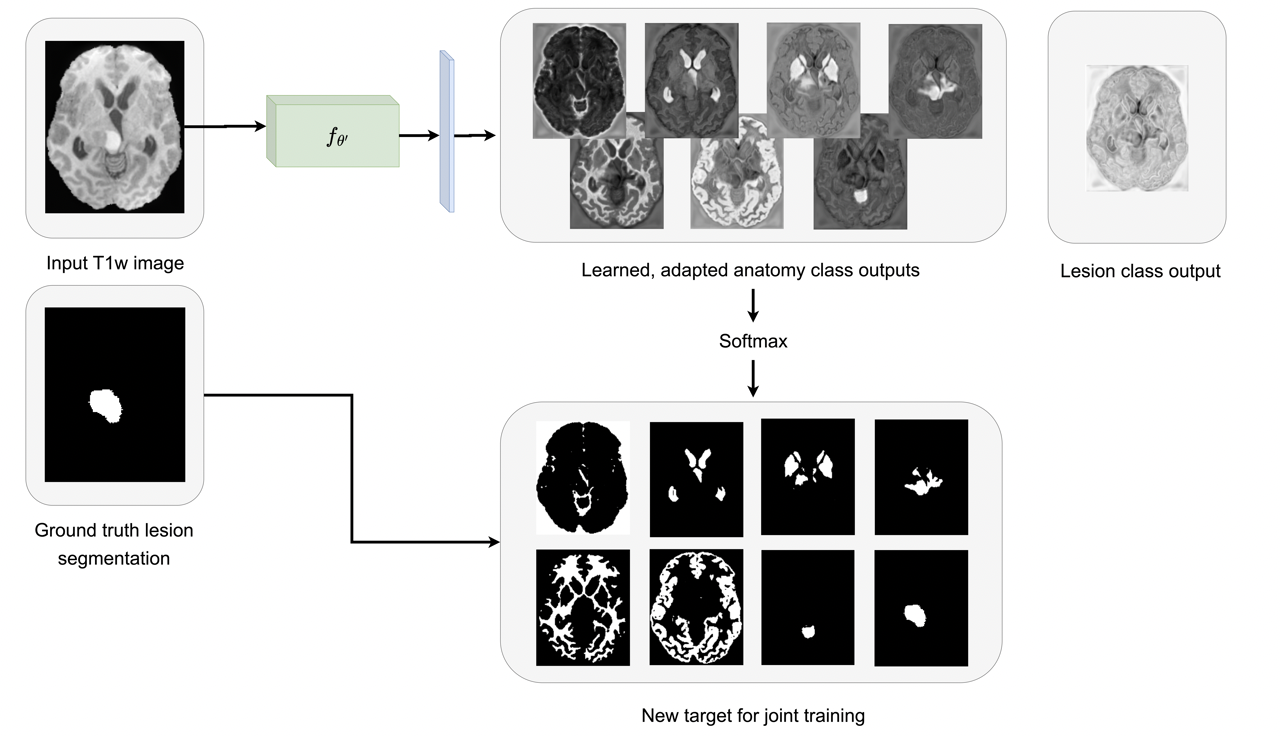}

    \caption{This figure demonstrates the way we utilize images with no ground truth anatomical segmentations in joint training. We use the network obtained after training the second stage, meta co-training. For each input T1w image, we obtain the adapted segmentation output for anatomical classes. For the lesion class, we have no information coming from this branch $f_\theta$. Then we combine the logits for anatomy classes after softmax operation, with the ground truth lesion segmentation as an additional channel. We use this probability array for the 8 classes as the target to calculate Dice loss with the outputs $\hat{y}$ in the next and final stage of joint training.  }
    \label{fig:pseudolabel}
\end{figure}

We explain the pseudolabel generation for joint training graphically in Figure \ref{fig:pseudolabel}.

\section{Varying lesion segmentation quality}

We demonstrate the effect on having worse than expected lesion segmentations on the adaptation steps of meta training. Since we have trained the meta co-training model to segment anatomical structures only, we can evaluate the anatomical structure segmentation with different lesion masks.

We remove some patches from the predicted lesion masks in testing to see this effect, namely we evaluate cases where we have half of the pixels of the lesion and the quarter of the pixels of the lesion in the testing lesion mask.

\begin{figure}
    \centering
    \includegraphics[width=0.4\linewidth]{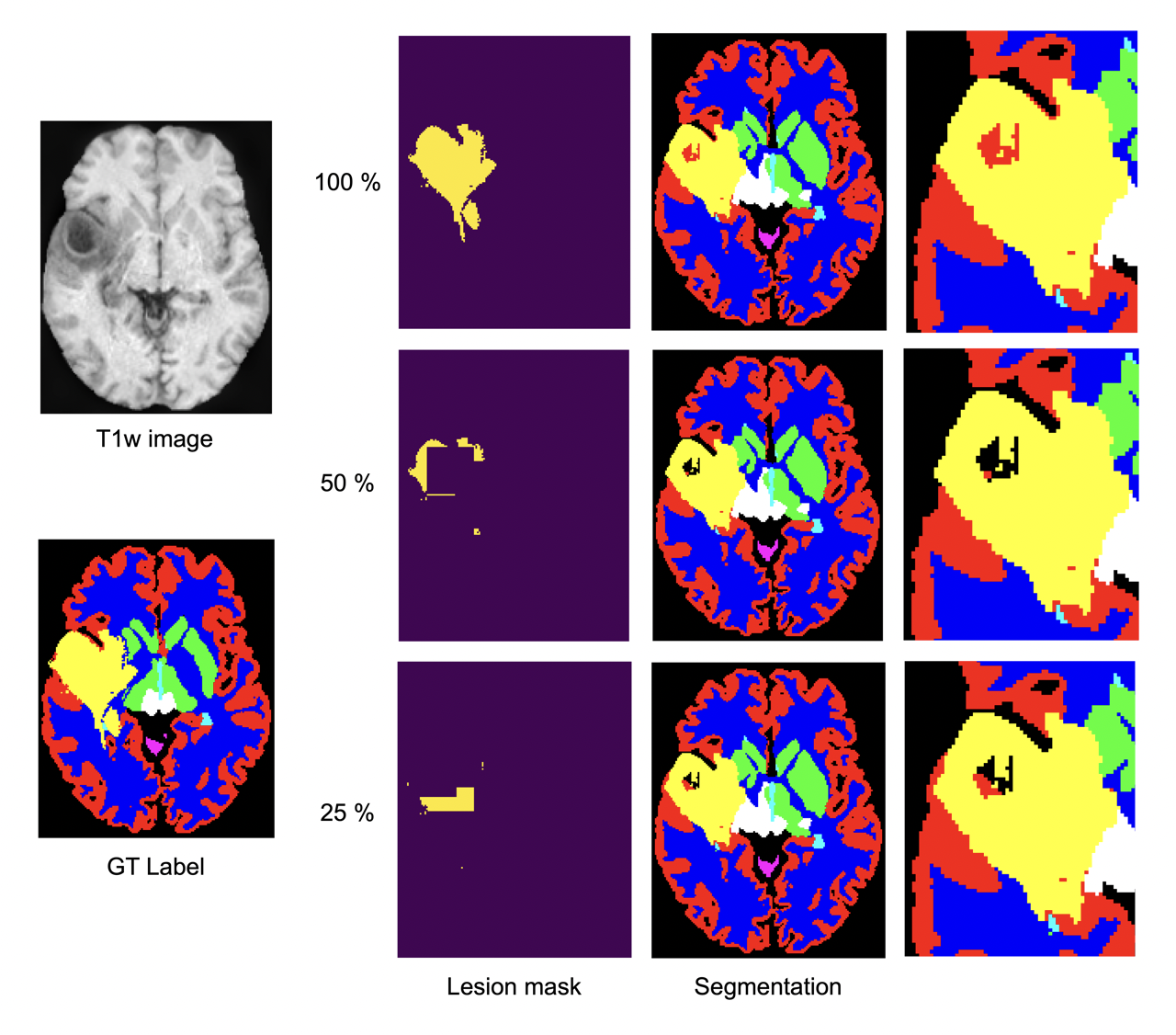}
    \caption{Example slice to demonstrate the effect of used lesion mask in testing after meta co-training stage. From top the bottom, we remove 25 percent of the lesion in the 3D mask randomly. The changes are minimal but can be observed in the gray matter border in the last column.}
    \label{fig:lesionmasks}
\end{figure}

We show the quantitative results on table below in Table \ref{tab:lesionmasks}, and a visual example in Figure \ref{fig:lesionmasks}. We see that the segmentation of the anatomical structure does not change drastically, but decrease slightly as the success of lesion segmentation decreases. It should also be noted that we add the predicted lesion segmentations here on top of the anatomical segmentation output from the meta co-training stage, which would not be available normally, thus the errors under those predicted lesion pixels are ignored.

From the visual results, we can see that the border of gray matter can be missegmented if the lesion is on the border of the brain, which creates a larger error. Overall, it is hard to evaluate since this is not the end-output of our intended model (as we demonstrated in the main results that joint training helps increase the scores for segmentation), and the error of the lesions would be further propagated, but from the results of our experiments the effects on the adaptation steps are minimal. Finally, we assume that we have the labeled lesion datasets available in our setting.

\begin{table*}[h!]{\label{tab:lesionmasks}}
\centering
\begin{tabular}{l|c|c|c|c|c|c|c|}
\cline{2-8}
 &{\textbf{Gray m.}}  	&	{\textbf{Basal g.}} 	&	{\textbf{White m.}}  	&	{\textbf{Tumor}} 	&	{\textbf{Vent.}} 	&	{\textbf{Cer.}} 	&	{\textbf{B. stem}} \\ \hline
  \multicolumn{1}{|l|}{\textbf{Full}} 	&	 {0.8733}&	{0.7890}  	&	 {0.9133}  &	 0.8844   	&	 {0.8729} 	&	 0.9576  	&	 0.8453\\ \hline
  \multicolumn{1}{|l|}{\textbf{Half}}&	 0.8728 &	 0.7892  &	0.9137  	&	 0.8844 	&	 0.8733   	&	 0.9572   	&	 {0.8445}   \\ \hline
  \multicolumn{1}{|l|}{\textbf{Quarter}} 	&	  {0.8725}   	&	 {0.7886}  &	 {0.9137}  &	 {0.8844} 	&	 {0.8722}   	&	 {0.9569}  &	 {0.8441}  \\ \hline
  \multicolumn{8}{c}{}
\end{tabular}
\caption{Dice scores for different classes in the meta co-training stage, varying the input test lesion mask used in masked test adaptation steps. Full denotes using the prediction of the lesion model, half denotes removing half of the predicted lesion pixels with 10x10x10 patches, and quarter denotes having quarter of the lesion pixels removed the same way.  }

\end{table*}

\begin{figure}
    \centering
    \includegraphics[width=0.7\linewidth]{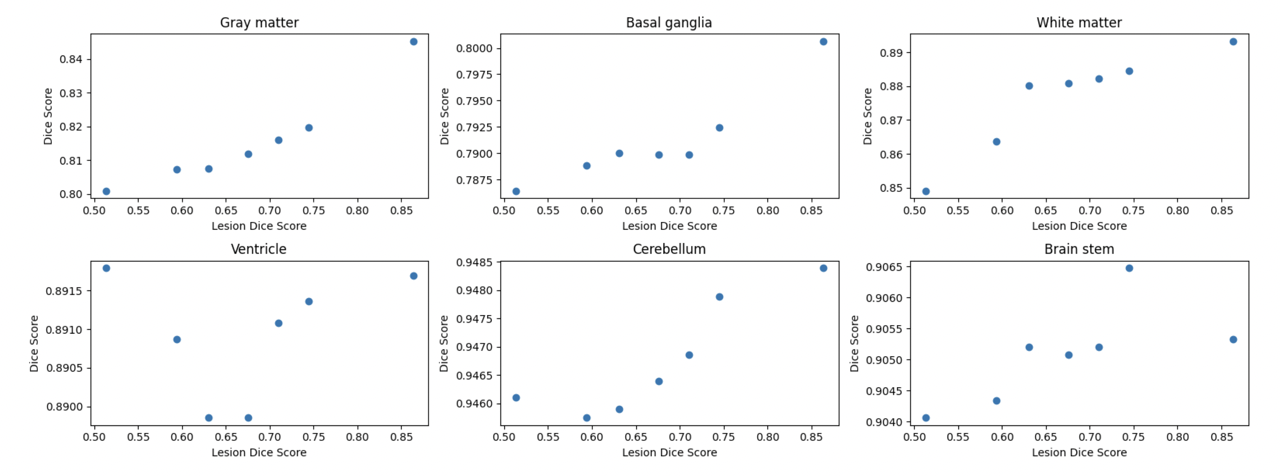}
    \caption{Dice score changes of the anatomical structures with respect to varying lesion segmentation success. One subject from the test set is shown, and the lesion segmentation features are varied by inputting augmented FLAIR sequence images (adding different amounts of noise).}
    \label{fig:lesionmasks-2}
\end{figure}

We additionally experiment with anatomy segmentation results with varying lesion segmentation performance. To evaluate this, we perform augmentations on the FLAIR images, resulting in worse segmentations than normal with the trained joint model. Then we use these lesion masks to test the joint model and get the whole segmentations. We show for one subject, the trend of Dice scores in Figure \ref{fig:lesionmasks-2}. We see that with better lesion masks, we obtain better segmentations in most of the classes overall, but the rate of change is slow with respect to the lesion Dice score. 

\newpage
\section{Lesion Masks in Joint Training}

In the joint training step, we create the randomized pixel values to calculate consistency, using the ground truth label map that belongs to that training image from BraTS dataset.
Since in the inference we do not have access to the ground truth masks, we use the predictions from the FLAIR model $f_\phi$ to change the values for randomization.
We experimented with using predicted lesion masks in the training as well, to create a better match for the expectation of the predicted mask in the test setting.

We share the difference in results on the test set in Table \ref{tab:traininglesion}.
The model trained with the predicted lesions tend to learn the lesion segmentation less accurately, leading to undersegmented lesions in the output, which results in a slightly worse performance on classes closer to the lesions (first three columns: gray matter, basal ganglia and white matter).

\begin{table*}[h!]{\label{tab:traininglesion}}
\centering
\begin{tabular}{l|c|c|c|c|c|c|c|}
\cline{2-8}
 &{\textbf{Gray m.}}  	&	{\textbf{Basal g.}} 	&	{\textbf{White m.}}  	&	{\textbf{Tumor}} 	&	{\textbf{Vent.}} 	&	{\textbf{Cer.}} 	&	{\textbf{B. stem}} \\ \hline
  {\multirow{2}{*}{\textbf{Ground truth}}} 	&	 {0.882}&	{0.790}  	&	 {0.921}  &	 0.881   	&	 {0.879} 	&	 0.958	&	 0.845\\
 \textbf{lesions}&  \small{(0.026)}&	\small{(0.046)} 	&	\small{(0.017)}  	&	\small{(0.064)} 	&	\small{(0.034)} 	&	 \small{(0.006)} &	 \small{(0.041)}\\ \hline
 {\multirow{2}{*}{\textbf{Predicted}}} 	&	 {0.880}&	{0.786}  	&	 {0.917}  &	 0.859   	&	 {0.878} 	&	 0.958	&	 0.841\\
 \textbf{lesions}&  \small{(0.027)}&	\small{(0.054)} 	&	\small{(0.017)}  	&	\small{(0.099)} 	&	\small{(0.034)} 	&	 \small{(0.007)} &	 \small{(0.044)}\\ \hline
\end{tabular}
\caption{Results of the joint model, trained with the ground truth label masks of the training images, and the predicted lesion masks from the FLAIR model. Average Dice score over 12 test subjects are shown on top, with the standard deviations on bottom.}

\end{table*}

\end{document}